\documentclass[twocolumn]{svjour3}          
\smartqed  
\usepackage{graphicx}

\usepackage{array}
\usepackage{amsmath}
\usepackage{amsfonts}
\usepackage{amssymb}
\usepackage{booktabs} 
\usepackage{graphicx}
\usepackage[hidelinks, pdfborder={0 0 0}]{hyperref}
\usepackage{natbib}
\usepackage[normalem]{ulem} 
\graphicspath{{./Figure/}}
	
\usepackage{color}
\usepackage{latexsym}
\usepackage{lscape}
\usepackage{natbib}

\newtheorem{assumption}[theorem]{Assumption}

\newcommand{\ignore}[1]{}

\newcommand{\Eqn}[1]{\ensuremath{\begin{array}{cccccccccc}#1\end{array}}}

\newcommand{\z}{{\mathbf z}}

\renewcommand{\epsilon}{\varepsilon}
{\begingroup
 \renewcommand{\theenumi}{\alph{enumi}}
 
 \begin{enumerate}}
 {\end{enumerate}\endgroup}

\newcommand{\II}{{\mathcal I}}

\usepackage{epsfig}
\usepackage{epstopdf}
\usepackage[usenames,dvipsnames]{xcolor}    

\newcommand{\mbF}{\mathbf{F}}

\newcommand{\mbx}{\mathbf{x}}
\newcommand{\mbu}{\mathbf{u}}

\newcommand{\subfigimg}[3][,]{%
  \setbox1=\hbox{\includegraphics[#1]{#3}}
  \leavevmode\rlap{\usebox1}
  \rlap{\hspace*{-5pt}\raisebox{\dimexpr\ht1-2\baselineskip}{#2}}
  \phantom{\usebox1}
}

\begin{document}

	\title{Variational and phase response analysis for limit cycles with hard boundaries, with applications to neuromechanical control problems}
	
	\titlerunning{Variational and phase response analysis of a neuromechanical control model}        
	
	\author{Yangyang~Wang \and Jeffrey P.~Gill \and Hillel J.~Chiel \and Peter J.~Thomas}
	
	
	\institute{Y.~Wang \at 
		Department of Mathematics, The University of Iowa, Iowa City, IA 52242, USA\\
		\email{yangyang-wang@uiowa.edu}
		\and J.P.~Gill \at 
		Department of Biology, Case Western Reserve University, Cleveland, OH 44106, USA\\
		\email{jpg18@case.edu}
		\and H.J.~Chiel \at
		Departments of Biology; Neurosciences; and Biomedical Engineering, Case Western Reserve University, Cleveland, OH 44106, USA
		\email{hjc@case.edu}
		\and P.J.~Thomas \at 
		Departments of Mathematics, Applied Mathematics, and Statistics; Biology; Cognitive Science; Data and Computer Science; and Electrical, Control and Systems Engineering, Case Western Reserve University, Cleveland, OH 44106, USA
		\email{pjthomas@case.edu}
	}
	\date{Received: /Accepted: }                        
	\maketitle
	
	\begin{abstract}
	Motor systems show an overall robustness, but because they are highly nonlinear, understanding how they achieve robustness is difficult. In many rhythmic systems, robustness against perturbations involves response of both the shape and the timing of the trajectory. This makes the study of robustness even more challenging. 
	
	To understand how a motor system produces robust behaviors in a variable environment, we consider a neuromechanical model of motor patterns in the feeding apparatus of the marine mollusk \textit{Aplysia californica} \citep{shaw2015,lyttle2017}. We established in \citep{WGCT2021} the tools for studying combined shape and timing responses of limit cycle systems under sustained perturbations and here apply them to study robustness of the neuromechanical model against increased mechanical load during swallowing. Interestingly, we discover that nonlinear biomechanical properties confer resilience by immediately increasing resistance to applied loads. In contrast, the effect of changed sensory feedback signal is significantly delayed by the firing rates' hard boundary properties. 
	Our analysis suggests that sensory feedback contributes to robustness in swallowing primarily by shifting the timing of neural activation involved in the power stroke of the motor cycle (retraction).
	This effect enables the system to generate stronger retractor muscle forces to compensate for the increased load, and hence achieve strong robustness. 
	
	The approaches that we are applying to understanding a neuromechanical model in \textit{Aplysia}, and the results that we have obtained, are likely to provide insights into the function of  other motor systems that encounter changing mechanical loads and hard boundaries, both due to mechanical and neuronal firing properties.
	
\end{abstract}

	\section{Introduction}
In many animals, motor control 
involves neural oscillatory circuits that can produce rhythmic patterns of neural activity without receiving rhythmic inputs (central pattern generators (CPGs)), force generation by muscles, and  interactions between the body and environment. 
Moreover, sensory feedback from the peripheral nervous system is known to modulate the rhythms of the electrical signals in CPGs and therefore facilitate adaptive behavior. 
	
	Motor systems show an overall robustness, but because they are highly nonlinear, understanding how they achieve robustness due to their different components is difficult. 
	To understand how animals produce robust behavior in a variable environment, \cite{shaw2015} and \cite{lyttle2017} developed a neuromechanical model of triphasic motor patterns to describe the feeding behavior of the marine mollusk \textit{Aplysia californica}. 
	Like many rhythmic motor systems,  feeding in \textit{Aplysia} involves two distinct phases of movement: a \emph{power stroke} during which the musculature engages with the substrate (the seaweed) against which it exerts a force to advance its goal (ingestion of seaweed), and a \emph{recovery stage} during which the motor system disengages from the substrate to reposition itself, in preparation for beginning the next power stroke.
	Similarly, in legged locomotion, the stance phase corresponds to the power stroke and the swing phase corresponds to the recovery stage.

\begin{figure*}[t] 
		\centering
		\includegraphics[width=0.9\textwidth]{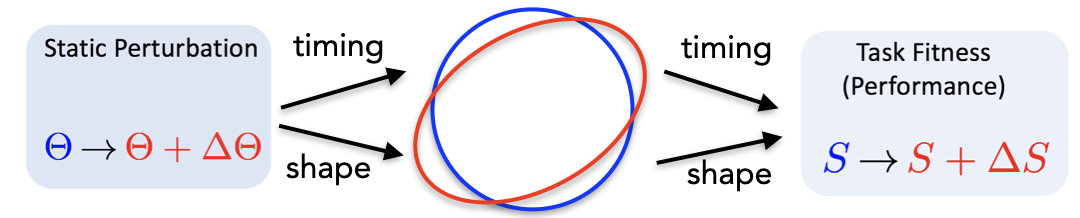}
		\caption{A sustained change in parameter $\Theta$ in a dynamical system $\dot{\mbx}=\mbF(\mbx,\Theta)$ producing a limit cycle trajectory typically causes changes in both the timing and shape of the trajectory, which may both influence the performance $S$ of the limit cycle system.}
		\label{fig:timing-shape} 
	\end{figure*}

	Also, like many rhythmic motor systems, feeding in \textit{Aplysia} involves a closed-loop system, which integrates biomechanics and sensory feedback, and exhibits a stable limit cycle solution.
	    It has been conjectured that sensory feedback plays a crucial role in creating robust behavior by extending or truncating specific phases of the motor pattern (\cite{lyttle2017}, \S 3.1).
	To test this hypothesis, we applied small mechanical perturbations as well as parametric perturbations to the sensory feedback pathways in the coupled neuromechanical system.
	It was shown in \cite{lyttle2017} that a sustained increase in mechanical load leads to changes in both shape and timing of the limit cycle solution: the system generates stronger retractor muscle force for a longer time in response to the increased load.
	Qualitatively similar effects have been observed during \textit{in vivo} experiments in \textit{Aplysia} \citep{Gill2020}.
	In general, we expect that applying parametric changes to CPG-based motor systems leads to changes in both the shape and timing of the resulting limit cycle behavior (Fig.~\ref{fig:timing-shape}).
	
	In \textit{Aplysia}, the increased duration (timing) and increased force (shape) have opposite effects on the task-fitness, measured as seaweed consumption per unit time. 
	Strengthening the retractor force pulls in more food with each cycle, which increases fitness, whereas lengthening the cycle time decreases fitness. 
	Together these effects approximately cancel, making the system robust against increased load. 
	This type of ``stronger-and-longer" response may occur generically in other motor systems. Thus, in this paper, we seek to understand the roles of sensory feedback and biomechanics in enhancing robustness. 
	To this end, we apply recently developed tools from variational analysis \citep{WGCT2021} to quantitatively study changes in shape and timing of a limit cycle under static perturbations.
	
	In the first part of the present paper (cf.~\S\ref{subsec:f-var}),  we apply the classical tools of forward variational analysis  to the model introduced by Shaw, Lyttle, Gill and coauthors in \citep{shaw2015,lyttle2017} (denoted as the Shaw-Lyttle-Gill or SLG model for brevity) to arrive at the following insights:
	
	\begin{itemize}
		
		\item Nonlinear biomechanical properties confer resilience by immediately increasing resistance to applied loads, on timescales much faster than neural responses.
	
		\item The main effect of sensory feedback is to shift the timing of retraction neural pool deactivation; in parallel, firing rate saturation effectively censors sensory feedback during specific movement subintervals.
		
		
	\end{itemize}    
	
	While the forward-in-time variational analysis is illuminating and allows us to explain in detail the robustness mechanism, it is still incomplete. Over time, the original and perturbed cycle will become increasingly out of phase due to the timing changes under sustained perturbations. Hence the shape displacements estimated from the forward variational analysis will become less and less accurate over time. 
	This difficulty is not limited to models of feeding in \textit{Aplysia californica}.
	For example, if we were to compare the gaits of two subjects walking on treadmills with slightly different speeds, although the ratio of stance and swing may be the functionally important aspect, this quantity is difficult to assess directly without putting the two movements on a common footing by comparing them using a common time scale.
	
	Thus, in order to compare perturbed and unperturbed motions with greater accuracy, in the remainder of the paper (cf.~\S\ref{subsec:isrc} and following) we show how to extend the local-in-time variational analysis to a global analysis by applying the \textit{infinitesimal shape response curve} (ISRC) analysis and \textit{local timing response curve} (LTRC) analysis developed in \cite{WGCT2021}.
	We review these methods in \S\ref{sec:methods}.
	This time-rescaled analysis accounts for both global timing sensitivity (through the \textit{infinitesimal phase response curve}, IPRC), as well as local timing sensitivity (through LTRC) by rescaling time to take into account local differences in the effects of parametric variation. 
	It yields a more accurate and self-consistent description of the oscillator trajectory's changing shape in response to parametric perturbations and helps complete the picture by providing a complementary perspective. 
	Specifically, our time-rescaled analysis provides additional insights, specifically that 
	\begin{itemize}
	
	    \item Increasing the applied load on the system increases the duty cycle of the neuron pool responsible for retraction, in the sense that the retraction neuron pool is activated for a larger \emph{percentage} of the closed phase of the cycle.
	    (The \emph{closed phase} of the trajectory occurs while the animal's radula-odontophore, or grasper, is closed on the seaweed, and encompasses the power stroke.)
	   	This effect ultimately results in more seaweed being consumed, despite increased force opposing ingestion. 

	   	\item We are able to derive the multidimensional \emph{infinitesimal phase response curve} (IPRC) despite the presence of nonsmooth dynamics in the system; we identify the mechanical component of the IPRC as the one that contributes most to robustness, and note that its contribution arises from the ``power stroke" segment of the motor cycle.

	   	\item We derive an analytical expression for the robustness to the mechanical perturbation
	   	that decomposes naturally into a sum of two terms, one capturing the effect of the perturbation on the shape of the trajectory, and the other capturing the effect on the timing; this result provides a \emph{quantitative} analysis of robustness that confirms the \emph{qualitative} insights described previously in the literature.

		\item In addition to sensory feedback and intrinsic biomechanical properties, robustness against changes in applied load can arise from coordinated changes of multiple parameters such as the gain of sensory feedback and muscle stiffness.
	
	\end{itemize}


	
	The dynamics of the SLG model \citep{lyttle2017} is given by
	\begin{eqnarray}\label{eq:Model}
	\Eqn{\frac{da_0}{dt}&=& (a_0(1-a_0-\gamma a_1)+\mu+\varepsilon_0(x_r-\xi_0)\sigma_0)/\tau_a \vspace{.05in}\\
		\frac{da_1}{dt}&=& (a_1(1-a_1-\gamma a_2)+\mu +\varepsilon_1(x_r-\xi_1)\sigma_1)/\tau_a\vspace{.05in}\\
		\frac{da_2}{dt}&=& (a_2(1-a_2-\gamma a_0)+\mu +\varepsilon_2(x_r-\xi_2)\sigma_2)/\tau_a\vspace{.05in}\\
		\frac{d u_0}{dt}&=& ((a_0+a_1)u_{\rm max}-u_0)/\tau_m\vspace{.05in}\\
		\frac{d u_1}{dt}&=& (a_2 u_{\rm max}-u_1)/\tau_m\vspace{.05in}\\
		\frac{dx_r}{dt}&=& (F_{\rm musc}(u_0,u_1,x_r) +rF_{\text{sw}})/b_r\\
		\vspace{1mm}}
	\end{eqnarray}
	This system incorporates firing rates of three neuron populations, corresponding to the ``protraction-open" ($a_0$), ``protraction-closed" ($a_1$), and ``retraction" phase ($a_2$). 
	Note that when a nerve cell ceases firing because of inhibition, its firing rate will be held at zero until the  balance of inhibition and excitation allow firing activity to resume. 
	Hence, we supplement model \eqref{eq:Model} with 
	three hard boundaries introduced by the requirement that the firing rates $a_0,a_1, a_2$ must be nonnegative:
	\[
	\Sigma_0=\{a_0=0\}, \Sigma_1= \{a_1=0\}, \Sigma_2=\{a_2=0\}.
	\]
	During the limit cycle, when a neural variable $a_i$ changes from positive to 0, we call that the $a_i$ \emph{landing point}; when it changes from 0 to positive, we call that the $a_i$ \emph{liftoff point}.
	The fact that the trajectory is non-smooth at the landing and liftoff points will play an important role in the analysis to follow.
	
	This model also consists of a simplified version of the mechanics of the feeding apparatus: a grasper that can open or close ($x_r$), a muscle that can protract the grasper to reach the food ($u_0$) and another muscle that can retract the grasper to pull the food back into its mouth ($u_1$). The net force exerted by the muscles is given by the sum of the two muscle forces 
	\footnotesize
	\begin{eqnarray}\label{eq:fmusc}
	F_{\rm musc}(u_0,u_1,x_r) &=&F_{\rm musc,pro}+F_{\rm musc,ret}\\&=&k_0\phi\left(\frac{c_0-x_r}{w_0}\right)u_0+k_1\phi\left(\frac{c_1-x_r}{w_1}\right)u_1\nonumber
	\end{eqnarray}
	\normalsize
	where 
	\[
	\phi(x)=-\frac{3\sqrt{3}}{2}x(x-1)(x+1)
	\]
	is the effective length-tension curve for muscle forces, $c_i$, $w_i$ and $k_i$ denote the mechanical properties of each muscle. 
	
	$F_{\text{sw}}$ represents the external force applied to the seaweed, which can only be felt by the grasper when it is closed on the food ($a_1+a_2>0.5$), during which $r=1$. When the grasper is open ($a_1+a_2\leq 0.5$), $r=0$; that is, the grasper moves independently of the seaweed. This condition leads to a transversal crossing boundary; that is, the open/closing boundary given by
	\[
	\Sigma_{o/c}=\{a_1+a_2=0.5\}.
	\]
    Values for model parameters and initial conditions are given in Table \ref{tab:parameters} and Table \ref{tab:initial_conditions} in the Appendix. For additional details on the biological assumptions motivating 
    the model, see \citep{shaw2015,lyttle2017}.
	
	In this paper, we are interested in the so-called heteroclinic mode in which the neural dynamics temporarily slow down when the sensory feedback overcomes the endogeneous neural excitation and forces the neural trajectory to slide along the hard boundary $a_i=0$. 
	Such temporary slowing down of neural variables allows the muscles, which evolve on slower timescales, to ``catch up'';  hence the seaweed can be swallowed and ingested successfully. 
	Following the terminology from \citep{WGCT2021}, we  call attracting periodic trajectories that experience sliding motions \emph{limit cycles with sliding components} (LCSCs).  
	
Using  classical sensitivity analysis and our recently developed tools from variational analysis \citep{WGCT2021}, we show here that biomechanics and sensory feedback cooperatively support strong robustness by changing the timing and shape of the neuromechanical trajectory. 
While both sensory feedback and biomechanics respond immediately to the increased load, we find that the sensory feedback effect is initially censored while the neural activity is pinned against a hard boundary of neuronal firing.  
Thus the effect of the sensory feedback signal is significantly delayed relative to onset of the increased load. 
Our analysis suggests that sensory feedback mediates robustness mainly through shifting the timing of neural activation and specifically increasing the duty cycle of the retraction neural pool. 
This response allows the system to digest more seaweed despite the increased force opposing ingestion and hence achieve strong robustness. 
In addition to uncovering the mechanisms for robust motor control, our methods allow us to   quantify analytically the robustness of the model system to the mechanical perturbation. 
Finally, although we focus on the \textit{Aplysia californica} feeding system as our working example, our methods should extend naturally to a broad range of motor systems.

	Our paper is organized as follows. We present our analysis and main results in \S\ref{sec:results}.
	Methods that we use to understand the robustness in the \textit{Aplysia} neuromechanical model are presented and reviewed in \S\ref{sec:methods}. We discuss limitations and possible extensions of our approach in \S\ref{sec:discussion}.
	
	\section{Results}\label{sec:results}
	
	Recall that a sustained (parametric) perturbation often causes changes in both shape and timing of the neuromechanical trajectory solution of \eqref{eq:Model}. In this paper, we adopt methods developed in \citep{WGCT2021} for analyzing the joint variation of both shape and timing of limit cycles with sliding components under parametric perturbations. 
	
	\subsection{Forward Variational Analysis.}\label{subsec:f-var}
	
	We begin our analysis by investigating how the shape of the trajectory changes in response to a sustained increase in mechanical load $F_{\text{sw}}$. To a first approximation, the change in shape can be captured by classical sensitivity analysis (also called forward variational analysis)  which we review in \S\ref{sec:methods}.   

We apply a small static perturbation to the system \eqref{eq:Model} by increasing the model parameter $F_{\text{sw}}$ by $\varepsilon=0.02$: $F_{\text{sw}}\to F_{\text{sw}}+\varepsilon$, and comparing the new, perturbed limit cycle trajectory $\gamma_\epsilon$ to the original, unperturbed limit cycle trajectory $\gamma_0$, beginning at the start of the grasper-closed phase (time 0 in Figure~\ref{fig:forward_var} panels A-D). 
	That is, we plot $\mbu\approx(\gamma_\epsilon(t)-\gamma_0(t))/\epsilon$; see \eqref{eq:forward_var_u}-\eqref{eq:forward_var_u0} for precise definitions.
	Note that $F_{\text{sw}}$ is multiplied by an indicator function that is only nonzero when the trajectory is in the grasper-closed phase. 
	Thus the perturbation is only present when the grasper is closed on the food. 
	
	The neural and biomechanical components of the unperturbed trajectory $\gamma(t)$ are shown by the solid curves in Figure~\ref{fig:forward_var}A and B, respectively. The perturbed trajectories $\gamma_\varepsilon(t)$ are indicated by the dashed lines. The gray shaded regions indicate phases when the grasper in the unperturbed system is closed.  
	With the perturbation (increased load), the transition from closing to opening is delayed; this transition is indicated by the magenta vertical line. 
	Figure \ref{fig:forward_var}C and D show the difference between the two trajectories per perturbation along the neural directions and along the biomechanical directions, respectively. 
	These curves can be approximated by the solutions to the forward variational equation \eqref{eq:forward_var} defined in \S \ref{sec:methods}. 
	The muscle forces $F_{\rm musc}(u_0, u_1, x_r)$  before and after the perturbation are shown by the gray curves in Figure \ref{fig:forward_var}B), and the difference between them is included as the gray curve in Figure \ref{fig:forward_var}D).   
	
	Figure \ref{fig:forward_var} yields several 
	insights about the roles of sensory feedback and biomechanics in robustness, which we discuss in detail below. 
	
	\subsubsection{Biomechanics confer resilience by immediately increasing resistance to the increased load.} 
	
\begin{figure*}[!htp] 
		\centering
		\includegraphics[width=0.9\textwidth]{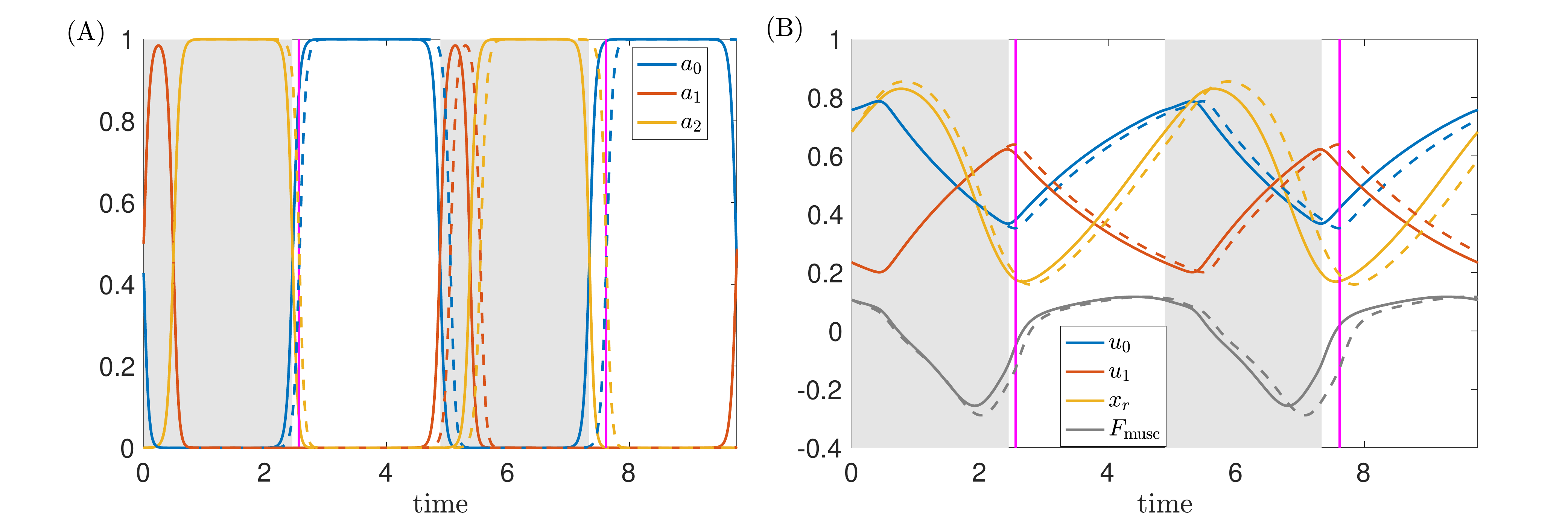}\\
		\includegraphics[width=0.9\textwidth]{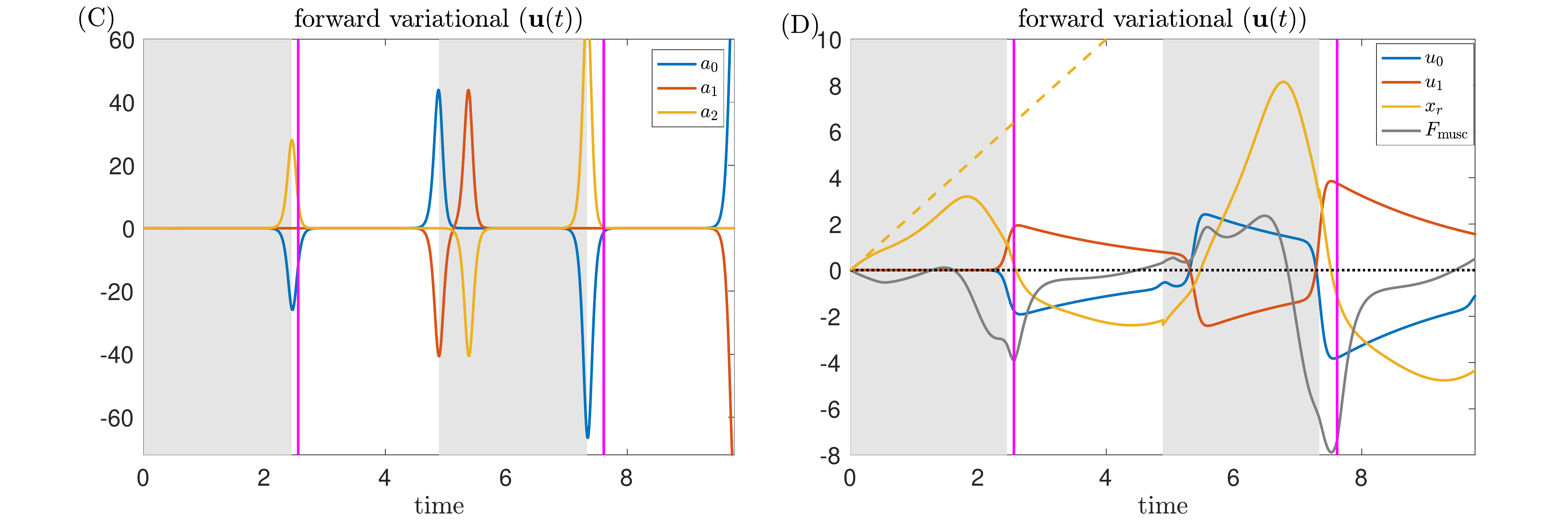}
		\caption{A small sustained perturbation is applied to the \textit{Aplysia} model \eqref{eq:Model} over the closing phase in which $F_{\text{sw}}\to F_{\text{sw}}+\varepsilon$ with perturbation $\varepsilon=0.02$.
			(A, B) Time series of the trajectory components for nominal force value $F_{\text{sw}}$ (solid) and perturbed force value (dashed) over two periods, aligned at the start of the closed phase.
			(C, D) The displacement solution, $\mbu(t)$, to the forward variational equation over two periods. 
			The gray curve in (D) denotes the displacement between the perturbed and unperturbed muscle forces $F_{\rm musc}$, shown as the gray curves in panel (B). 
			The yellow dashed line in panel (D) approximates the displacement in $x_r$ if the net muscle force $F_{\rm musc}$ did not change after perturbation. 
			The intervals during which the grasper without perturbation is closed on the food are indicated by the shaded regions.
			The vertical magenta lines indicate the times at which the grasper under perturbation switches from closed to open.
			The difference in periods and the delay in the grasper opening time both accumulate, making the
			comparison between the two trajectories invalid except for short times.
			(A) and (C) show trajectories and displacements along the neural directions, while (B) and (D) show trajectories and displacements along the mechanical directions. 
			The lines in panel (C) (resp., (D)) approximate the difference between the dashed and solid trajectories from panel (A) (resp., (B)) per perturbation.   }
		\label{fig:forward_var} 
	\end{figure*}

	Immediately following the perturbation of the mechanical load, we observed a positive displacement in the grasper position $x_r$ relative to the unperturbed trajectory (Figure \ref{fig:forward_var}D, yellow curve). 
	This displacement simply reflects the grasper 
	being pulled by a stronger force $F_{\text{sw}}+\varepsilon$. 
	If nothing other than the applied load $F_{\text{sw}}$ changes in the system, a linearized analysis suggests that the initial displacement of $x_r$ would approximately follow the yellow dashed line given by $\frac{1}{b_r}t$ (Figure \ref{fig:forward_var}D). 
	Nonetheless, while this line gives a good initial approximation, the true displacement in $x_r$ (yellow solid curve) quickly sags below the yellow dashed line over time. 
	This difference arises due to the negative displacement occurring in the muscle force $F_{\rm musc}(u_0, u_1,x_r)$ (Figure \ref{fig:forward_var}D, gray curve) which acts to overcome the increased load.  
	However, early in the retraction cycle, all other variables, including the muscle activation $u_0$ and $u_1$, show almost no displacement at all (see Figure \ref{fig:forward_var}C and D). 
	This observation suggests that long before the sensory feedback effect has time to act, the biomechanics may play an essential role in generating robust motor behavior, by providing an immediate, short-term resistance to increased load.

	
	Early in the retraction cycle, increasing the load stretches both the retractor and protractor muscles, and moves them down their length-tension curves. 
	As a result, both forces become weaker, but the magnitude of the protractor muscle force drops more quickly than the retractor muscle force (Figure \ref{fig:biomechanics}). 
	Thus, the retractor muscle force grows relative to the opposing protractor muscle force.
	This shift endows the system with a built-in resilience, in that increasing seaweed force opposing ingestion automatically (i.e., without changes in neural activation) engages a larger resisting muscle force long before neural pools or muscles show differential activation. 
	This is a new insight beyond the previous ``longer-stronger" hypothesis \citep{shaw2015,lyttle2017}.
	
	\subsubsection{Sensory feedback effects are largely delayed by the firing rate hard boundary properties.}  
	
	Changes in $x_r$ due to the increased load will immediately propagate to the neural variables ($a_0, a_1, a_2$) through the sensory feedback $\varepsilon_i(x_r-\xi_i)\sigma_i)/\tau_a$, and hence should affect the neural variables without any lag. 
	However, no significant displacement of the neural variables is observed until nearly the end of the retraction cycle (Figure \ref{fig:forward_var}C). 
	In other words, while the sensory feedback itself immediately responds to the increased load, the \emph{effect} of the changed sensory feedback signal is not manifest until much later in the retraction cycle, when the protraction-open neuron pool is released from inhibition along its hard boundary and starts to fire (Figure \ref{fig:forward_var}). 
	Hence, the nonsmooth 
	hard boundary conditions on neuronal firing rates significantly delay the effect of sensory feedback, and create intervals during which neurons are insensitive to sensory feedback.
	This effect is a concrete example of \emph{differential penetrance} \citep{Ye2006,Cullins2015}.

	\subsubsection{Sensory feedback contributes by shifting the \emph{timing} of neural activation.}
	
\begin{figure}[!htp] 
		\centering
		\includegraphics[width=0.5\textwidth]{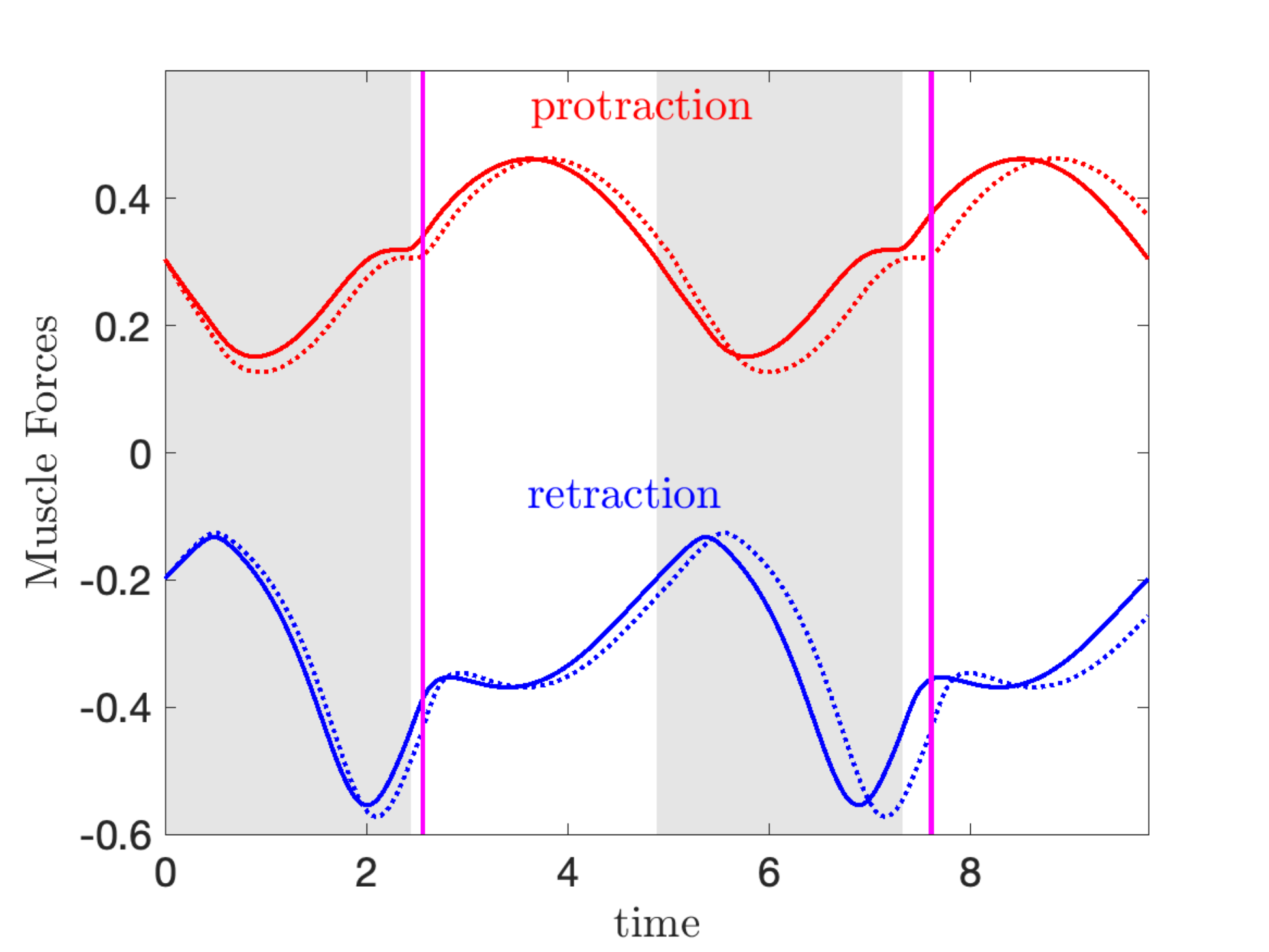}
		\caption{The time series of the perturbed (dashed) and unperturbed (solid) protractor muscle forces $F_{\rm musc,pro}$ (red) and retractor muscle forces $F_{\rm musc,ret}$ (blue) over two periods. The gray shaded regions and the magenta lines have the same meanings as in Figure \ref{fig:forward_var}.}
		\label{fig:biomechanics} 
	\end{figure}

	Due to the hard boundary effects, the displacements of the neural variables appear near the end of the closed phase, when the protraction-open neuron pool $a_0$ lifts off from its hard boundary, and the retraction-closed neuron pool $a_2$ deactivates to stop firing (Figure \ref{fig:forward_var}A and C). 
	A positive (resp., negative) displacement of $a_2$ (resp., $a_0$) indicates that $a_2$ deactivates (resp., $a_0$ activates) at a later time with the increased load, and hence the retraction-closed phase is prolonged. 
	Consequently, the retraction muscle activity will increase, because its stimulation by the retraction motor neuron is prolonged, allowing the slow retractor muscle to generate larger forces (Figure \ref{fig:biomechanics}D). 
	Similarly, we also observe a decreased protractor muscle activity, as the protraction neuron pool $a_0$ turns on at a later time. 
	This \emph{decrease} leads to a stronger retractor muscle force and a weaker protractor muscle force (Figure \ref{fig:biomechanics}). 
	Hence, a more negative net muscle force results (Figure \ref{fig:forward_var}D, gray curve), which corresponds to a stronger resisting force pulling the seaweed towards the jaw to swallow the food. 
	To summarize, the main effect of sensory feedback that contributes to robustness is prolonging the retraction phase to 
    confer on the system a resilience in that increasing seaweed force opposing ingestion engages a larger resisting muscle force. 
    Thus, sensory feedback contributes to robustness primarily by shifting the \emph{timing} of neural activation, as opposed to the magnitude of neural activation. 
    Biologically, this distinction corresponds to affecting the timing of motor neuron burst onset or offset, rather than burst intensity.
    
	\subsection{Variational Analysis with Rescaled Time - ISRC.}\label{subsec:isrc}
	
	\begin{figure}
	    \centering
	    \includegraphics[width=0.4\textwidth]{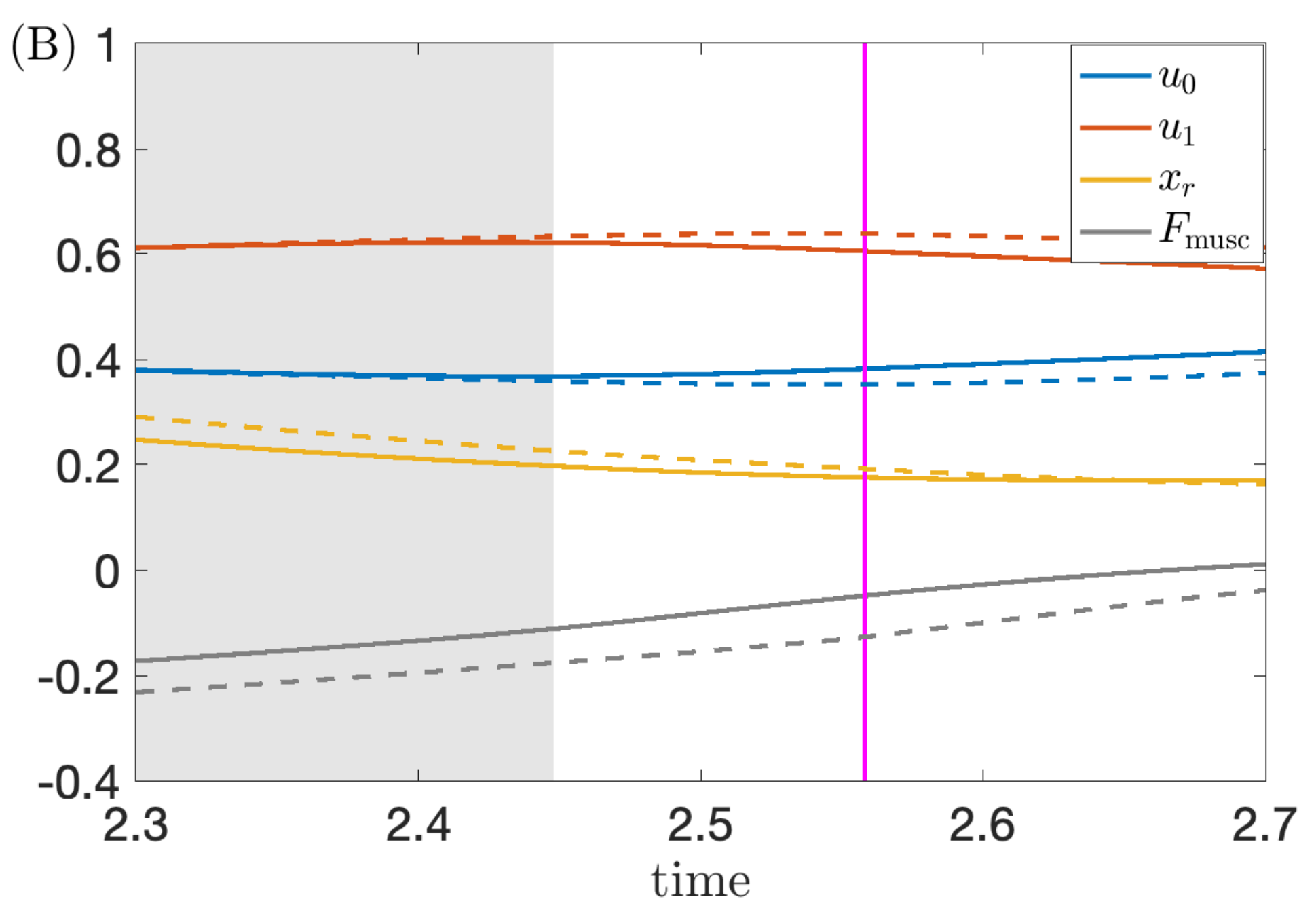}
	    \includegraphics[width=0.4\textwidth]{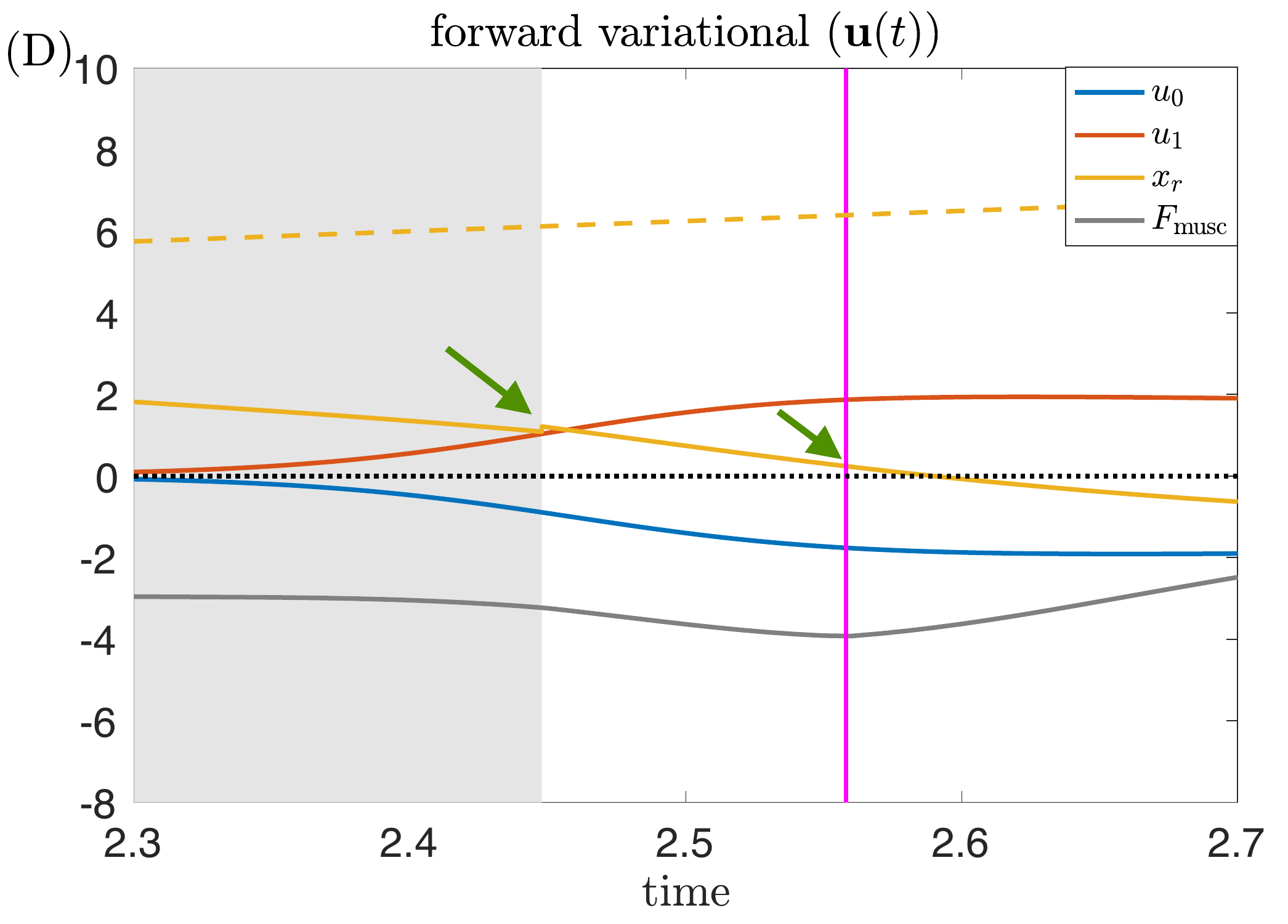}
	    \caption{Enlarged views of Figure \ref{fig:forward_var}B and D near the first transition from closed to open. Note the displacement of $x_r$ (i.e., the $x_r$ component of $\mathbf{u}(t)$ shown as the yellow solid line in (D)) is positive both at the closing time of the unperturbed system (see the green arrow near the grey/white boundary) and the perturbed system (see the green arrow near the magenta line).}
	    \label{fig:forward_var_detail}
	\end{figure}
	
	Under the  forward-in-time analysis, the grasper of the perturbed system lags behind the grasper of the unperturbed system 
	throughout the closing phase; yet the \emph{net} seaweed movement was measured to be greater for the perturbed system \citep{lyttle2017}. 
	Fig.~\ref{fig:forward_var_detail} shows an expanded view of the perturbed and unperturbed systems' grasper position (Fig.~\ref{fig:forward_var_detail}B) and the linearized difference produced by the variational equation (Fig.~\ref{fig:forward_var_detail}D).  
	As this detailed view shows, at the time when the unperturbed system transitions from closed to open (gray-white boundary) the unperturbed grasper position is \emph{more retracted} than the perturbed grasper position at the coincident time point.
	Similarly, the grasper component of the variational equation is \emph{positive} at the gray-white boundary.  
	Furthermore, at the time when the perturbed system transitions from closed to open (magenta line) the perturbed system continues to be less retracted than the unperturbed system.  
	Thus, whether we compare the systems at the perturbed or unperturbed opening time, the perturbed grasper is ``further behind". Yet the overall effect in the perturbed system is a larger net intake of seaweed per cycle.

	This apparent contradiction underscores the need to extend the perturbation analysis beyond the standard forward-in-time variational analysis.  
	In particular, 
	if one cycle is slower than another, then while the local perturbation analysis can explain the cause-and-effect relations a short time into the future, they cannot account for the net effect around a cycle in a self-consistent way.
	Over time, the displacements between the two trajectories grow, and the linearized approximation becomes invalid except at short times (cf.~\cite{JS07}). 
	Hence, unless time is rescaled to take into account the difference in cycle period,
	comparing the components of the original and perturbed cycles will become less and less meaningful.
	
\begin{figure*}[!htp] 
		\centering
		\includegraphics[width=0.9\textwidth]{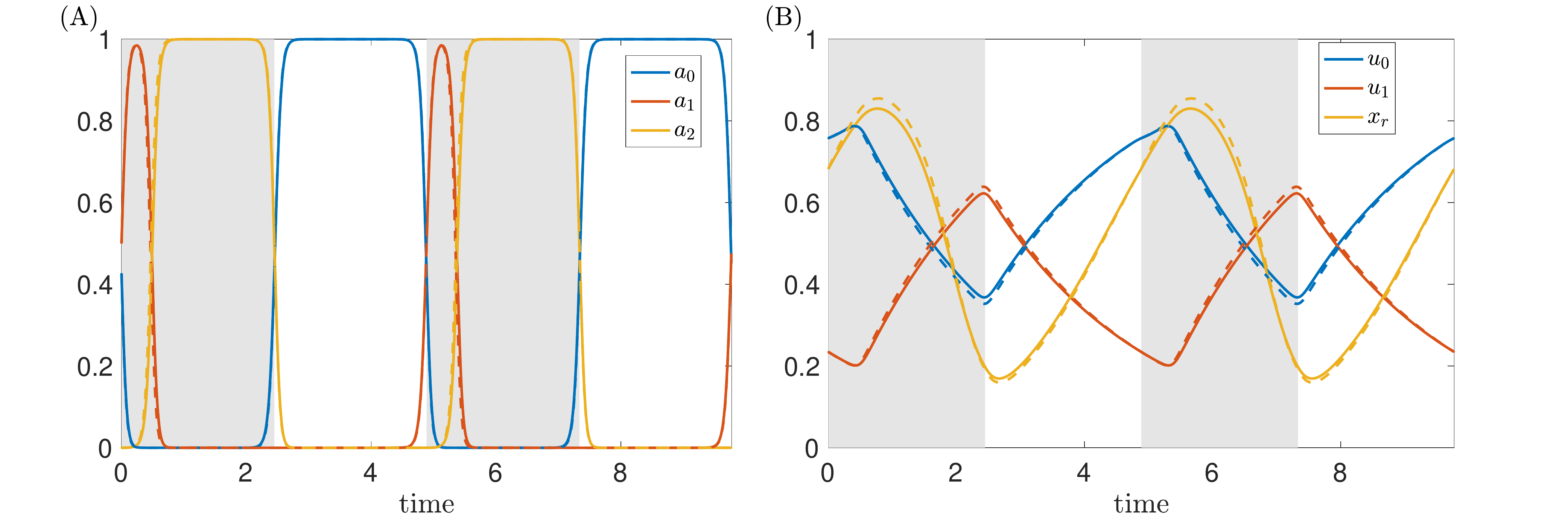}\\
		\includegraphics[width=0.9\textwidth]{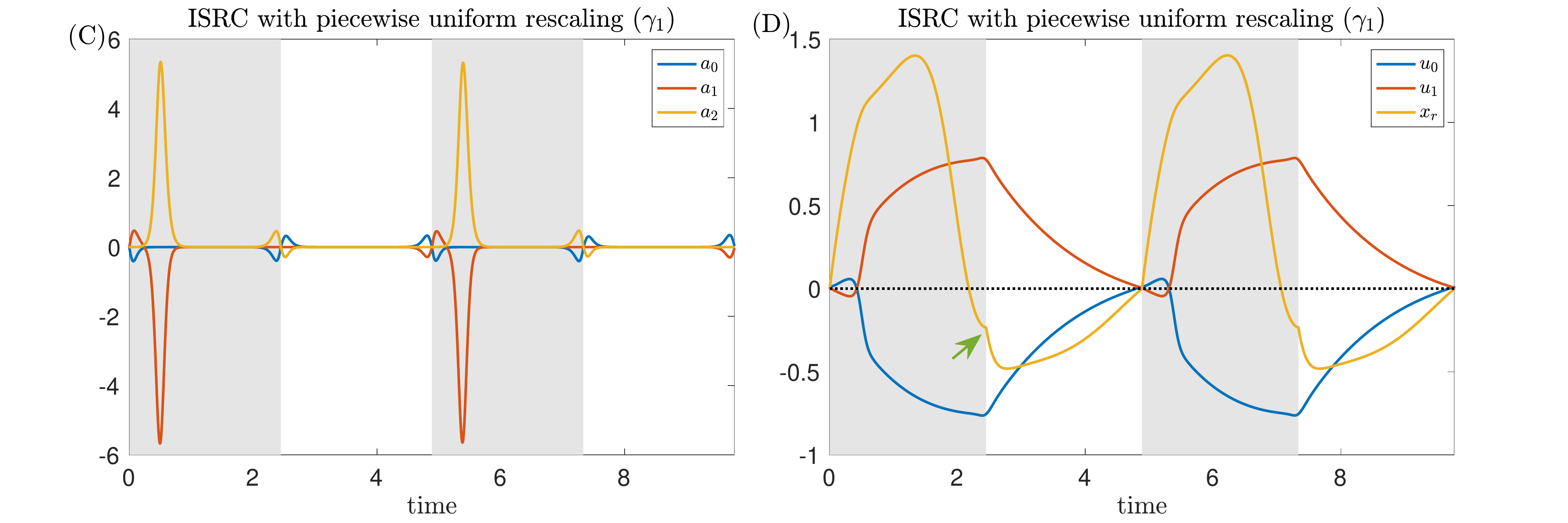}
		\caption{Variational analysis with piecewise uniform time rescaling.
			The same sustained perturbation as in Figure \ref{fig:forward_var} is applied to the \textit{Aplysia} model \eqref{eq:Model}.
			(A, B) Time series of the perturbed (dashed) and unperturbed solutions (solid).
			Here piecewise uniform rescaling is applied so the closing and opening events coincide.
			(C, D) The ISRC with piecewise uniform rescaling $\gamma(t)$ over two periods.
			Shaded regions have the same meanings as in Figure \ref{fig:forward_var}. Note the $x_r$ component of the ISRC is negative at the time of opening (see green arrow).
			With piecewise uniform rescaling, the variational approximation is consistent across multiple periods (c.f., Figure \ref{fig:forward_var}).}
		\label{fig:isrc_pw} 
	\end{figure*}

	To overcome this difficulty, we extend the local-in-time variational analysis to a global analysis by rescaling time so the unperturbed closing and opening events coincide with those after perturbations, respectively. We do so by applying the \textit{infinitesimal shape response curve} (ISRC) analysis and the \textit{local timing response curve} (LTRC) \citep{WGCT2021}, which we review in \S\ref{sec:methods}. 
	This method yields a more accurate and self-consistent description of the oscillator trajectory's changing shape in response to parametric perturbations (see Figure \ref{fig:isrc_pw}).
	We show that the combination of the ISRC and the LTRC gives a sensitivity analysis of an oscillator to sustained perturbations within any given region (e.g., protraction or retraction cycle, opening or closing phase) and provides a self-contained framework for analytically quantifying and understanding robustness to perturbations.
	
	We write $\gamma_1$ for the linear shift in the limit cycle shape in response to the static perturbation $F_{\text{sw}}\to F_{\text{sw}}+\varepsilon$, that is: 
	\[\gamma_\epsilon(\tau_\varepsilon(t))=\gamma_0(t)+\epsilon\gamma_1(t)+O(\epsilon^2),\] 
	uniformly in time.
	Note that the time for the perturbed trajectory is rescaled to be $\tau_\varepsilon(t)$ to match the unperturbed time points.
	The linear shift $\gamma_1(t)$ is the so-called ISRC curve and satisfies a nonhomogeneous variational equation (see \S\ref{sec:methods}). 
	Compared with the forward variational equation, the ISRC equation has one additional nonhomogeneous term $\nu_1 F_0(\gamma_0(t))$ that arises from the time rescaling. 
	In this term, $\nu_1$ is determined by the choice of time rescaling $\tau_\varepsilon(t)$ and $F_0(\gamma_0(t))$ is the unperturbed vector field evaluated along the unperturbed limit cycle $\gamma_0(t)$ (see \S\ref{sec:methods} for details). 
	
	Since the perturbation is applied to the seaweed, it can only be felt by the system when the grasper is closed on the seaweed. 
	It is natural to expect that the segment at the closing phase has a different timing sensitivity than the segment at the opening phase. 
	We hence choose to rescale time differently in the two phases, using piecewise uniform rescaling when computing the ISRC. 
	This leads to a piecewise ISRC equation, where $\nu_1$ is piecewise constant. 
	It was shown in \citep{WGCT2021} that $\nu_1$ can be estimated from the LTRC analysis (see \S\ref{sec:methods}). 
	
	In Figure \ref{fig:isrc_pw}A and B, the time traces of variables along the unperturbed limit cycle are shown by the solid curves, whereas the perturbed limit cycle whose time has been rescaled to match the unperturbed time points as described above are indicated by the dashed curves. With the piecewise rescaling, the transitions between the closing and opening events of the perturbed and unperturbed systems now coincide. The relative displacements between the perturbed and unperturbed trajectories are approximately given by the piecewise ISRC $\gamma_1$ shown in Figure \ref{fig:isrc_pw}C and D. 
	In contrast to the forward variational analysis, in which the displacements grow over time, the piecewise ISRC curve is periodic, meaning we have achieved a self-consistent global description of the response of the limit cycle to increased load. 
	
	We now show that the apparent contradiction that we obtained from the forward variational analysis, i.e., that the grasper displacement at the end of the closing phase is positive (cf.~Fig.\ref{fig:forward_var_detail}), can now be resolved in the time-rescaled picture. 
	In response to the perturbation, the relative displacement of the grasper position (the $x_r$ component of $\gamma_{1}$, detoted as $\gamma_{1,x_r}$) initially increases (i.e., the grasper becomes more and more protracted due to the increased load) and reaches its peak at about $t=1.4$ (see Figure \ref{fig:isrc_pw}D, yellow curve). Then it starts to decrease and becomes negative at the time when the grasper opens. This means later in the retraction cycle, the perturbed grasper is less and less protracted than the unperturbed version and eventually become more retracted by the end of the closing phase (Fig.~\ref{fig:isrc_pw}D, green arrow). 
	In summary, the grasper perturbed by larger force begins ``behind" the unperturbed version, but catches up around 60\% of the way through the retraction phase (in relative time) and comes out ``ahead" by the time  both graspers open, consistent with having a larger net seaweed intake \citep{lyttle2017}.
 
	To understand what causes $\gamma_{1,x_r}$ to be negative despite its initial big rise, we consider the effect of the perturbation on the neural pool through sensory feedback. 
	In Figure \ref{fig:isrc_pw}C, we observe positive displacements in $\gamma_{1,a_2}$ (yellow curve) occurring both when the retraction neuron pool $a_2$ activates and when it deactivates. 
	These displacements indicate that with the increased load, the retraction neuron $a_2$ activates earlier and turns off later relative to the unperturbed $a_2$.
	In other words, increasing the applied load on the system increases the \emph{duty cycle} of the neuron pool involved in retraction, i.e., the retraction neuron pool is activated for a larger percentage of the total cycle. 
	As a result, the motor system recruits a larger retractor muscle force, as indicated by the positive displacement of the retractor muscle activation $u_1$ during the closing phase (Figure \ref{fig:isrc_pw}D, red curve). 
	A similar increase in motor recruitment in response to increased external load has been observed \textit{in vivo} \citep{Gill2020}.
	In the model, the stronger retraction force acts to impede the protraction of the grasper, and eventually pulls the grasper to a more retracted state.
	Thus the grasper displacement crosses zero and becomes negative at the end of the closing phase (Figure \ref{fig:isrc_pw}D, green arrow). 
	
		
	Note that there is no perturbation during the opening phase (Figure \ref{fig:isrc_pw}, white space). 
	During this phase of the cycle, displacements slowly decay and are nearly zero by the time the grasper closes on the food again.

	\subsection{Timing responses to sustained perturbations of $F_{\text{sw}}$.}\label{sec:iprc-aplysia}
	
	\subsubsection{Infinitesimal phase response curve.}\label{sec:results-iprc}
	
\begin{figure}[!htp]
		\begin{center}
 	\includegraphics[width=0.5\textwidth]{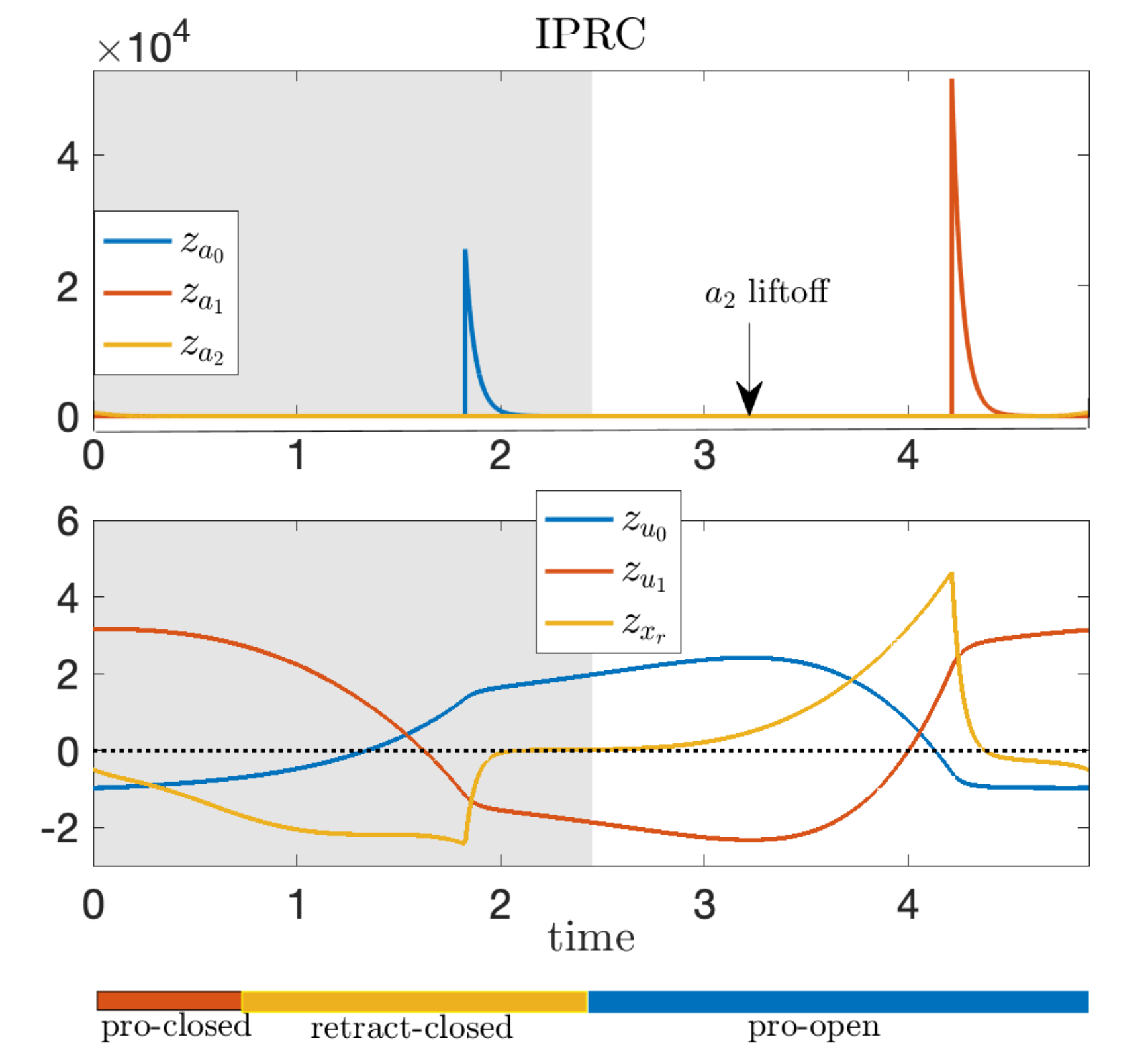}
		\end{center}
		\caption{\label{fig:prc-aplysia} IPRC for the \textit{Aplysia} model. Grey shaded region indicates the period when the radula/odontophore is closed. On the bottom, the red, yellow and blue rectangles denote the protraction-closed, retraction-closed and protraction-open phases, respectively. 
		The blue spike in the IPRC in the top panel occurs when the $a_0$ variable ``lifts off'' from the $a_0=0$ boundary; 
		the red spike occurs when the $a_1$ variable lifts off;
		the liftoff point for $a_2$ is indicated with an arrow.
		}
	\end{figure}

	To understand the timing response of system \eqref{eq:Model} to increased load, we perform an IPRC analysis. Figure \ref{fig:prc-aplysia} shows the time traces of the IPRC curve over one cycle. As before, the shaded region indicates the phase when the grasper is closed. 
		 
    The IPRC curves associated with biomechanical variables are shown in Figure \ref{fig:prc-aplysia}, lower panel. In particular, the timing sensitivity of system \eqref{eq:Model} to the increased load on the grasper ($F_{\text{sw}}\to F_{\text{sw}}+\varepsilon$) can be estimated using the IPRC along the $x_r$ direction, i.e., the yellow curve in the lower panel of Figure \ref{fig:prc-aplysia}. Since the perturbation only has effect during the closing phase, only the portion of $z_{x_r}$ over the shaded region is relevant. This portion is strictly negative. Therefore, in response to the increased load considered above, the system undergoes phase delay, and hence the total period is prolonged. This finding is consistent with earlier results on the sensory feedback effect obtained from the variational analysis (see \S\ref{subsec:isrc}). 
    
    The linear shift in period can be estimated by evaluating the integral  \[T_1:=\lim_{\varepsilon\to 0}\frac{T_{\varepsilon}-T_0}{\varepsilon}= -\int_{0}^{T_0} \z(t)^\intercal \frac{\partial F_\varepsilon(\gamma_0(t))}{\partial \varepsilon}\Big|_{\varepsilon=0} \,dt,\] 
    where $T_0$, $T_\varepsilon$ are the periods before and after perturbation $\varepsilon$ (see Section \ref{sec:methods}). For the perturbation on $F_{\text{sw}}$, the derivative $\frac{\partial F_\varepsilon(\gamma_0(t))}{\partial \varepsilon}$ equals $(\mathbf{0},\frac{1}{b_r})^\intercal$ over the grasper-closed region, and equals $\mathbf{0}$ during the grasper-open region, where the first $\mathbf{0}$ is a $5\times 1$ zero vector and the second $\mathbf{0}$ is a $6\times 1$ vector. 
    It then follows that 
    \begin{equation}\label{eq:T1-fsw}
     T_1=-\int_{\Theta_{\rm close}} z_{x_r}(t)/b_r\, dt
     \end{equation}
     where $\Theta_{\rm close}$ denotes the grasper-closed phase.  
    
    Other IPRC curves in Figure \ref{fig:prc-aplysia} indicate the timing sensitivity of the model to other perturbations and lead to several useful insights as well as testable predictions. For example, 
    \begin{itemize}
        \item The IPRC curves are continuous except at the liftoff points (Figure \ref{fig:prc-aplysia} top panel, blue and red spikes). While all three neural variables go through liftoff points, there is no large spike in $z_{a_2}$ (yellow curve). The absence of a yellow spike and the fact that the red spike is larger than the blue spike, imply that the system has the highest timing sensitivity to perturbing $a_1$ and intermediate timing sensitivity to $a_0$, both of which are significantly higher than the sensitivity to $a_2$ perturbations.  
        
        \item Excitatory inputs to neural populations lead to phase advance and hence shorten the total period, because the IPRC curves associated with neural variables are mostly positive (Figure \ref{fig:prc-aplysia} top panel). 
        
        \item  Most of the time the system is not sensitive to neural perturbations, but there also exist sensitive regions when the trajectory is not restricted to the hard boundaries (e.g., Figure \ref{fig:prc-aplysia} top panel, blue and red spikes). For instance, the system has high timing sensitivity to perturbations of $a_0$ late in the closing phase and to perturbations of $a_1$ late in the opening phase, whereas sensory inputs are largely ignored early in the opening phase.  
         
        \item Increasing the protractor muscle activation $u_0$ causes a phase delay early in the closing phase and late in the opening phase, and a phase advance otherwise. In contrast, increasing the retraction muscle activation $u_1$ causes a phase advance early in the closing phase and late in the opening phase, and a phase delay otherwise.  Appendix \ref{sec:different-time-sensitivity} discusses why the system has different timing sensitivities to muscle perturbations. 
    \end{itemize}

   Although all three neural variables go through liftoff points, there is no large yellow spike in $z_{a_2}$ (see Figure \ref{fig:prc-aplysia}). To understand this, we note that before $a_0$ (resp., $a_1$) lifts off its hard boundary, there exists no inhibition from other neurons except for inhibitory sensory feedback. However, when $a_2$ lifts off at around $t\approx 3.2$, it still experiences inhibition from $a_0$ (see Figure \ref{fig:isrc_pw}A). 
   In other words, there are two inhibitory effects pressing neurons $a_2$ down to the hard boundary, but only one inhibitory effect acting on the other two neuron populations. As a result, while there is a discontinuous jump of the IPRC curve corresponding to $a_2$ at the liftoff point, it remains small as the other inhibition is still present. 
    
\subsubsection{Local timing response curve.}\label{sec:results-LTRC}
     
\begin{figure}[!htp]
\begin{center}
\includegraphics[width=0.5\textwidth]{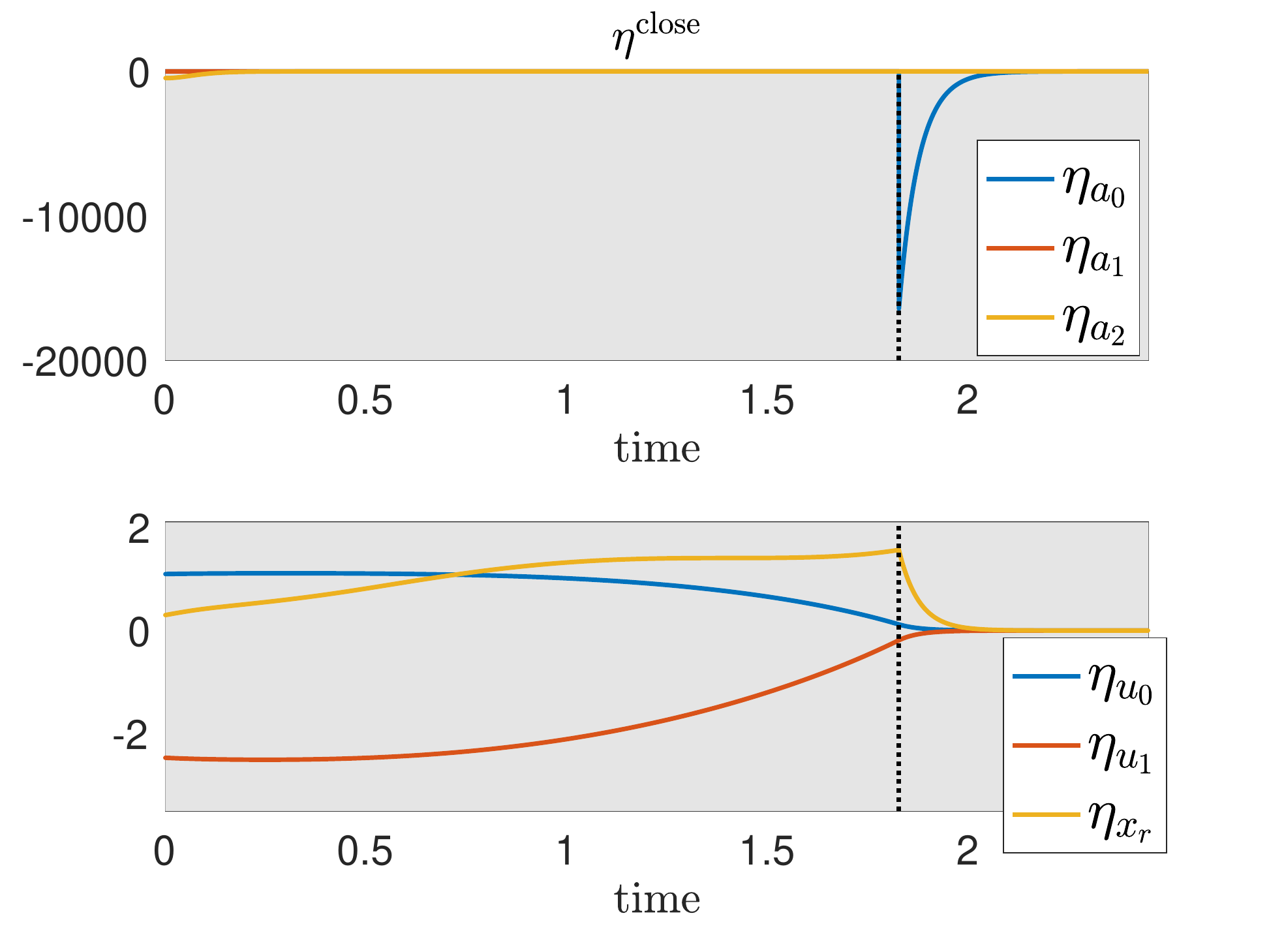}
\end{center}
\caption{\label{fig:LTRC-close} Time series of the LTRC $\eta^{\rm close}$ over the closing phase. The liftoff point on $a_0=0$ coincides with the spike in $\eta_{a_0}$ (blue curve, top panel). The cusp where $\eta_{x_r}$ changes from increasing to decreasing (intersection of yellow and vertical black dashed curves in the bottom panel) also occurs at the liftoff point for $a_0$.}
\end{figure}

  While the IPRC is a powerful tool for understanding the global timing sensitivity of an oscillator to sustained perturbations, it does not give local timing sensitivities, which, however, are needed for computing the ISRC curve as discussed above. We hence adopt the local timing response curve (LTRC) method developed in \cite{WGCT2021} and reviewed in \S\ref{sec:methods}. To illustrate this method, we show the LTRC associated with the closing phase and denote it as $\eta^{\rm close}$ (see Figure \ref{fig:LTRC-close}). 
Although the LTRC $\eta^{\rm close}$ is defined throughout the full domain, estimating the effect of the perturbation within the closing region only requires evaluating the LTRC in this region. Figure \ref{fig:LTRC-close} shows the time series of $\eta^{\rm close}$ for the model in the closing region, obtained by numerically integrating the adjoint equation backward in time with the initial condition of $\eta^{\rm close}$ given by its value when the grasper switches from closing to opening. Note that $\eta_{x_r}$, the yellow curve in Figure \ref{fig:LTRC-close} lower panel, remains positive over the closing phase. This implies that the increased load on seaweed prolongs the time remaining in the closing region; that is, the increased load prolongs the total closing time. The relative shift in the closing time caused by the increased load can also be estimated by integrating the LTRC (see Section \ref{sec:methods}). 
	
In addition, Figure \ref{fig:LTRC-close} implies that strengthening the protractor muscle activation $u_0$ during the closing phase prolongs the total closing time, whereas increasing the retraction muscle activation $u_1$ decreases the total closing time.  Similarly, we can compute the LTRC over other phases, such as the retraction phase, in order to estimate local timing sensitivities of the system in other regions. 

Finally we note an interesting feature in $\eta^{\rm close}$: there is an abrupt change in $\eta_{x_r}$ at the $a_0=0$ liftoff point  (Figure \ref{fig:LTRC-close} bottom panel, dashed vertical line). 
To understand this behavior, note that an instantaneous perturbation of $x_r$ directly propagates to neural pools through sensory feedback. 
While all three neural pools are affected by this mechanical perturbation, the neural components of $\eta^{\rm close}$ are zero most of the time except when the trajectory lifts off from the $a_0=0$ constraint (Figure \ref{fig:LTRC-close} top panel, blue spike). 
This observation implies that the system has a high local sensitivity to $a_0$ during the blue spike, whereas the sensitivity to $a_1$ and $a_2$ are significantly smaller than unity at all times. 
Thus, to understand the effect of perturbing $x_r$ on the local timing, it is sufficient to focus on $\eta_{a_0}$ and examine how $a_0$ reacts to perturbing $x_r$. 

Similar to the forward variational analysis, perturbing $x_r$ delays the activation of $a_0$, i.e., $a_0$ lifts off from $a_0=0$ at a later time. That is, the displacement in $a_0$ near the $a_0=0$ liftoff point is negative. Since $\eta_{a_0}$ is negative near the liftoff point, perturbing $x_r$ prolongs the total closed time (i.e., $\eta_{x_r}$ is positive during the closed phase).   

Next we address the cusp phenomena observed in $\eta_{x_r}$ (Figure \ref{fig:LTRC-close}, bottom panel, yellow curve). 
Note that perturbations arriving before the trajectory lifts off from $a_0=0$ delay the activation of $a_0$ by increasing the inhibition from its sensory feedback. 
Moreover, the closer the time of perturbation to the time of liftoff, the larger the delay on the activation of $a_0$. Such a larger delay leads to a greater increase of the total closed time due to perturbing $x_r$. 
Hence, before the liftoff time (Figure \ref{fig:LTRC-close}, bottom panel, vertical black dashed line), $\eta_{x_r}$ gradually increases. 
Once the trajectory has passed the liftoff point, perturbing $x_r$ delays the activation of $a_0$ by decreasing its sensory feedback, the effect of which now becomes excitatory. 
The size of this effect decays exponentially as the trajectory gradually leaves the boundary $a_0=0$. 
Thus, there is a cusp in the $\eta_{x_r}$ curve at the liftoff point, after which $\eta_{x_r}$ rapidly decreases.

\subsection{Robustness to static perturbations.}
	
In this section, we show how the robustness of the \textit{Aplysia} model \eqref{eq:Model}, the ability of the system to maintain its performance despite perturbations, can be quantified using the ISRC, IPRC and LTRC analysis. 

Following \citep{lyttle2017}, we quantify the performance or task fitness via the average seaweed intake rate 
\begin{equation}\label{eq:fitness}
    S_{\varepsilon} = \frac{-\Delta x_{r,\varepsilon}}{T_\varepsilon}
\end{equation}
where $\Delta x_{r,\varepsilon}$ is the net change in perturbed grasper position $x_{r,\varepsilon}$ during the grasper-closed phase and $T_\varepsilon$ is the perturbed period. Note that we assume the seaweed is moving together with the grasper when it is closed and not moving at all during the grasper-open component of the trajectory. Hence, $-\Delta x_{r,\varepsilon}$ denotes the total amount of seaweed consumed per cycle. 

Since the vector field $F_{\varepsilon}(\mbx)$ in system \eqref{eq:Model} is piecewise smooth in the coordinates $\mbx$ and smooth in the perturbation $\varepsilon$, it follows that the following expansion holds:
\[
\Delta x_{r,\varepsilon} = \Delta x_{r,0} + \varepsilon \Delta x_{r,1}+O(\varepsilon^2),
\]
where $\Delta x_{r,0}$ is the net change in the unperturbed grasper position during the grasper-closed component of the trajectory. 
Here, $\Delta x_{r,1}$ is approximately given by the net change of the $x_r$ component of the ISRC $\gamma_1$, which is denoted as $\gamma_{1,{x_r}}$ (see \S\ref{subsec:isrc}), over the grasper-closed phase. 
Suppose the grasper closes at $t^{\rm close}$ and opens at $t^{\rm open}$ over one cycle. 
It follows that $\Delta x_{r,1} =\gamma_{1,{x_r}}(t^{\rm open}) -  \gamma_{1,{x_r}}(t^{\rm close})$. 

\citep{lyttle2017} show that the robustness, i.e., the relative shift in the task fitness per relative change in perturbation, for small $\varepsilon$, can be written as   
\begin{align}\label{eq:robustness}
\text{Robustness}&=\frac{F_\text{sw}}{\varepsilon}\frac{S(\varepsilon)-S_0}{S_0}\\
\nonumber 
&=F_\text{sw}\left(\frac{\Delta x_{r,1}}{\Delta x_{r,0}}-\frac{T_1}{T_0}\right)+\mathcal{O}(\varepsilon),
\end{align}
as $\varepsilon\to 0$.
Recall that $T_0$ is the period of the unperturbed limit cycle and $T_1$ denotes the linear shift in period, which can be estimated using the IPRC (see \S\ref{sec:results-iprc}). 

In summary, the robustness formula can be decomposed into two parts, one involving changes in shape (in particular, the grasper position $x_r$) and the other involving the timing change. As discussed before, changes in shape can be estimated using the ISRC and the LTRC analysis, whereas the latter can be quantified using the IPRC. Below, we illustrate the quantification of the robustness by considering the perturbation to be the increase in the constant applied load $F_{\text{sw}} \to F_{\text{sw}}+\varepsilon$. 

The ISRC with or without timing rescaling corresponding to the perturbation on the applied load have already been computed and discussed in \S\ref{subsec:f-var} and \S\ref{subsec:isrc}. 
Note that $\Delta x_{r,1}$ in \eqref{eq:robustness} is the net change in the ISRC during the grasper-closed phase. 
Choosing the ISRC with rescaling based on the timing of the closing and opening events provides a more accurate estimate of $\Delta x_{r,1}$. 
Hence, we use the ISRC with piecewise rescaling to estimate $\Delta x_{r,1}$, which is the net change in $\gamma_{1,{x_r}}$ over the closing region per cycle (see the yellow curve over the shaded region in the lower right panel of Figure \ref{fig:isrc_pw} and the green arrow marking the difference at the end of the closing phase). 
Furthermore, the linear shift in the period $T_1$  can be estimated by \eqref{eq:T1-fsw} using the IPRC. 
	
\begin{figure}[!t]
\begin{center}
\includegraphics[width=8cm]{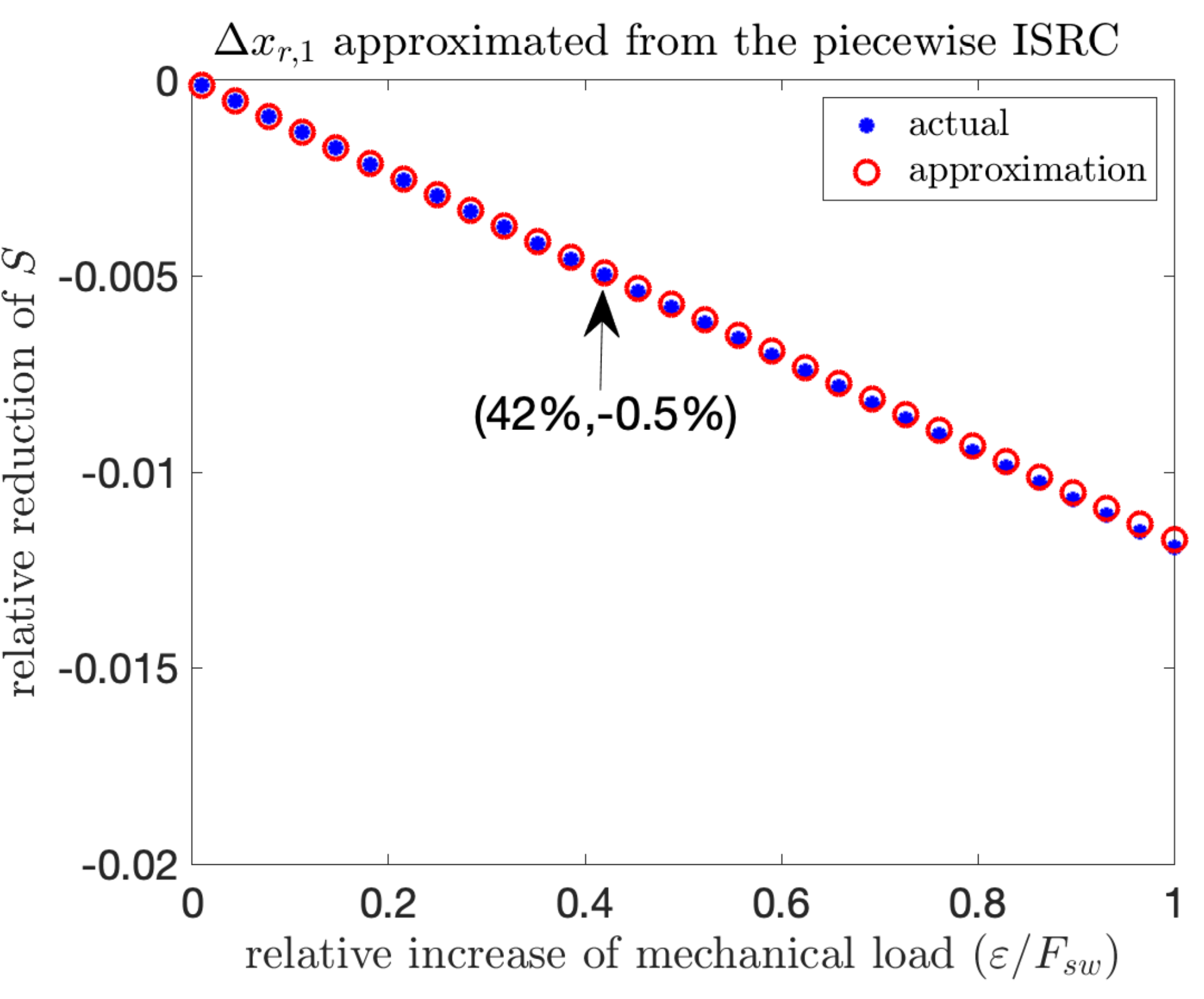}
\includegraphics[width=8cm]{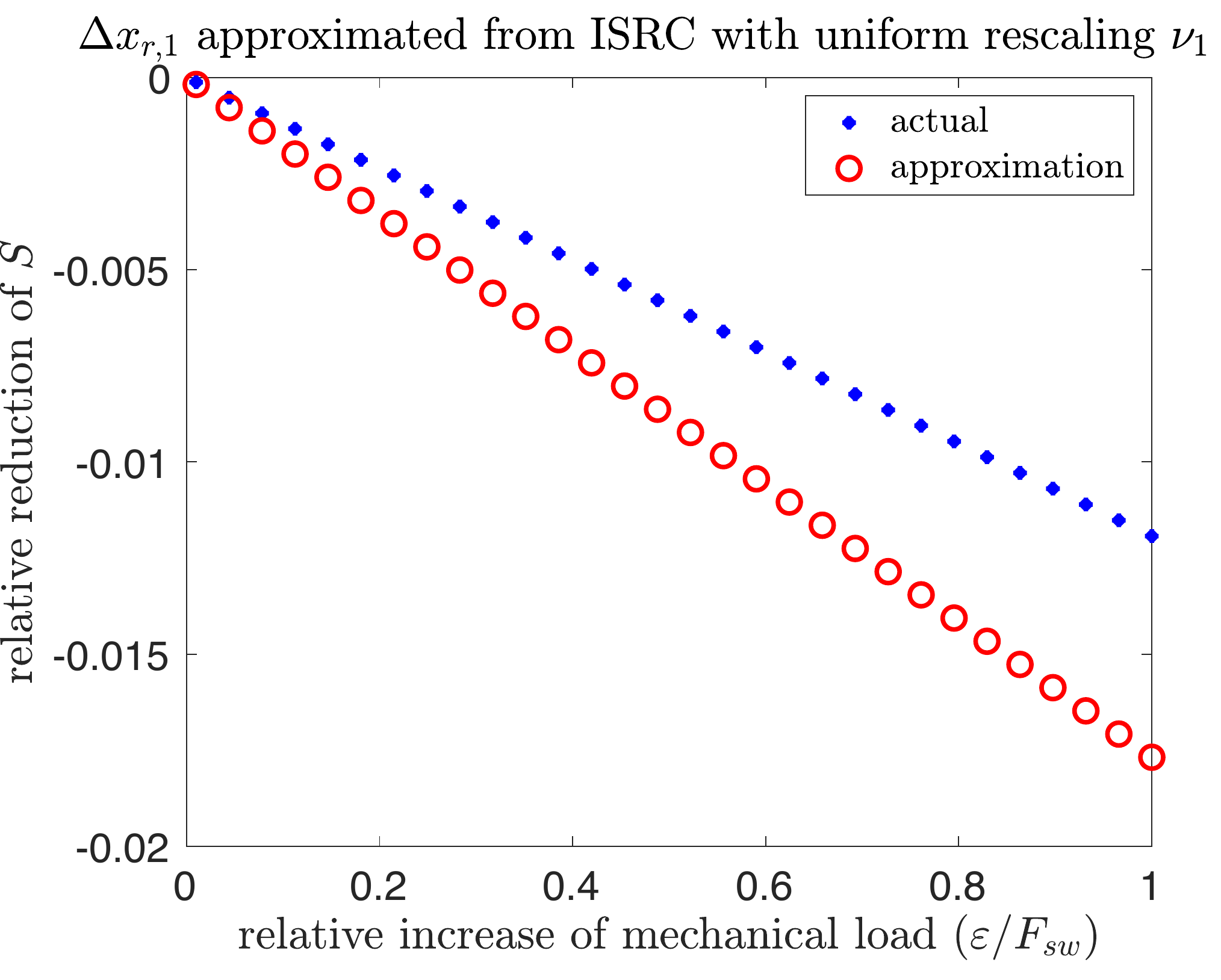}
\end{center}
\caption{\label{fig:robustness} Relative change in task fitness $(S(\varepsilon)-S_0)/S_0$ computed numerically (blue stars) versus those obtained analytically from the ISRC and the IPRC according to  formula \eqref{eq:robustness} (red circles), as the perturbation $\varepsilon$ on the seaweed load $F_{\text{sw}}$ varies. Without perturbation, the nominal applied load is $F_{\text{sw}}=0.01$. The approximation using the ISRC with different timing rescalings during the grasper-closed ($\nu_{1,\rm close}$) versus grasper-open phases ($\nu_{1,\rm open}$) estimated from the LTRC analysis matches the actual simulation (top panel), whereas the ISRC with uniform rescaling $\nu_1=T_1/T_0$ estimated from the IPRC no longer gives a good approximation (bottom panel).}
\end{figure}  

    From the above analysis, we obtain $\frac{\Delta x_{r,1}}{\Delta x_{r,0}}\approx 0.4806$ and $\frac{T_1}{T_0}\approx 1.6532$, both of which are positive and are consistent with the concept of an adaptive ``stronger and longer" change in the motor pattern in response to increased load. 
    It follows that the robustness is approximately $-1.1726\times 10^{-2}$. 
    (Note that the smaller this number is in magnitude, the more robust the system is.)    
    To the first order in $\varepsilon$, the relative change in the performance is then given by 
    $$\frac{S(\varepsilon)-S_0}{S_0}\approx -1.1726\times 10^{-2}(\varepsilon/F_\text{sw}),$$ 
    which is illustrated by the red circle as the perturbation size $\varepsilon$ varies (see Figure~\ref{fig:robustness}, top panel). 
    To see what this means, we take a data point on the line indicated by the arrow, i.e., $(0.42,-0.005)$. 
    Here $\varepsilon/F_{\rm sw}=0.42$ indicates a $42\%$ increase in load $F_{\text{sw}}$, which only causes a $0.5\%$ decrease in the task fitness, corresponding to a highly robust response.
    Here the ``stronger" effect (i.e., the first term in the robustness formula \eqref{eq:robustness} being positive) contributes to the robustness whereas the ``longer" effect (i.e., the second term in the robustness) reduces it. However, these two effects are not independent from each other: it is the longer retraction-closed time that allows the muscle to build up a stronger force, thereby contributing to a robust response.  
    
    We also compute the relative change in $S$ with respect to $\varepsilon$ using direct numerical simulations (see Figure~\ref{fig:robustness}, blue stars), which show good agreement with our analytical results. 
    In contrast, if we estimate $\Delta x_{r,1}$ using the ISRC with a uniform timing rescaling (see Figure \ref{fig:isrc_open_close}), the resulting estimated robustness becomes more negative and no longer gives an accurate approximation to the actual robustness (see Figure~\ref{fig:robustness}, bottom panel). 
    That is, the ISRC using different rescaling factors over the grasper-closed phase ($\nu_{1,\rm close})$ versus the grasper-open phase ($\nu_{1,\rm open}$), gives a much better approximation to the robustness than the ISRC based on a global timing rescaling $\nu_1=T_1/T_0$. 
    The fact that the $\nu_{1,\rm{open/close}}$ are obtained via the LTRC analysis highlights the contribution of this novel analytical tool.
    This observation demonstrates that for systems under certain circumstances (e.g, non-uniform perturbation as considered in system \eqref{eq:Model}), 
     the ISRC together with the LTRC greatly improves the accuracy of the robustness, compared to the ISRC with global timing analysis given by the IPRC.

		
		
		
		
	
\subsection{Sensitivity of robustness to other parameters.}
\label{ssec:other-param}

In general, the performance of motor control systems may be affected not only by external parameters, such as an applied load, but also be internal parameters, for instance describing the physical properties of the biomechanics or neural controllers.  
The variational tools used in the previous section to understand mechanisms of robustness to increases in applied load -- the IPRC, ISRC and LTRC -- can also give insights into the effects of changing internal model parameters. 
For instance, in the SLG model, appropriately varying strengths of protractor or retractor muscles can overcome effects of the increased mechanical load $F_{\text{sw}}\to F_{\text{sw}}+\varepsilon$. 
Because of the SLG model's relative simplicity, we can relate many of these changes to specific components of the fitness equation in detail.  
	

Below, we first consider how varying sensory feedback strengths can help restore the reduced seaweed intake rate due to increased applied load. Then we examine how changing the strengths of the protractor and retractor muscles affects robustness to applied loads.

	\subsubsection{Varying sensory feedback strengths.}
	
\begin{figure}[!htp]
		\begin{center}
		\includegraphics[width=7cm]{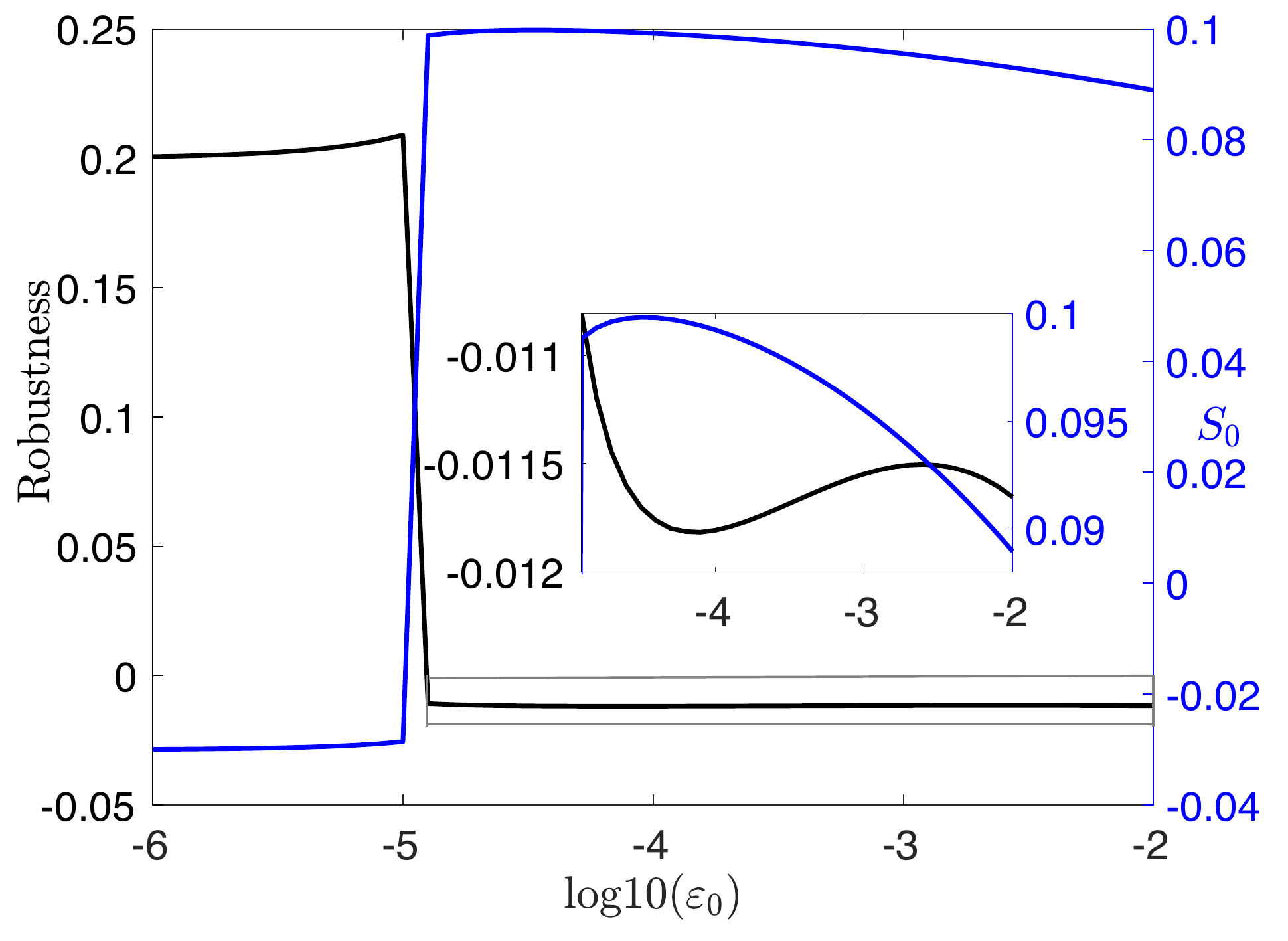}
		\includegraphics[width=7cm]{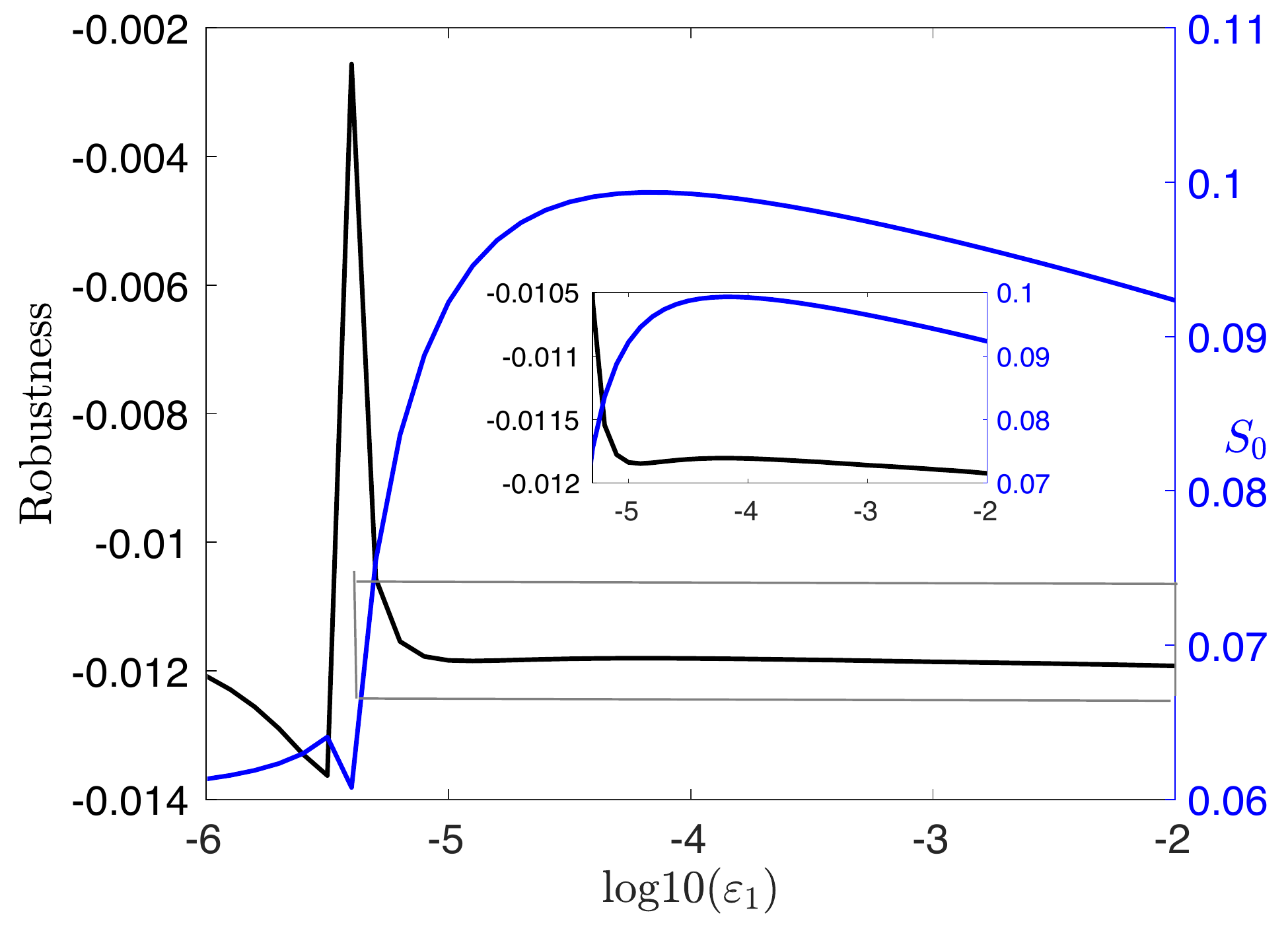}
		\includegraphics[width=7cm]{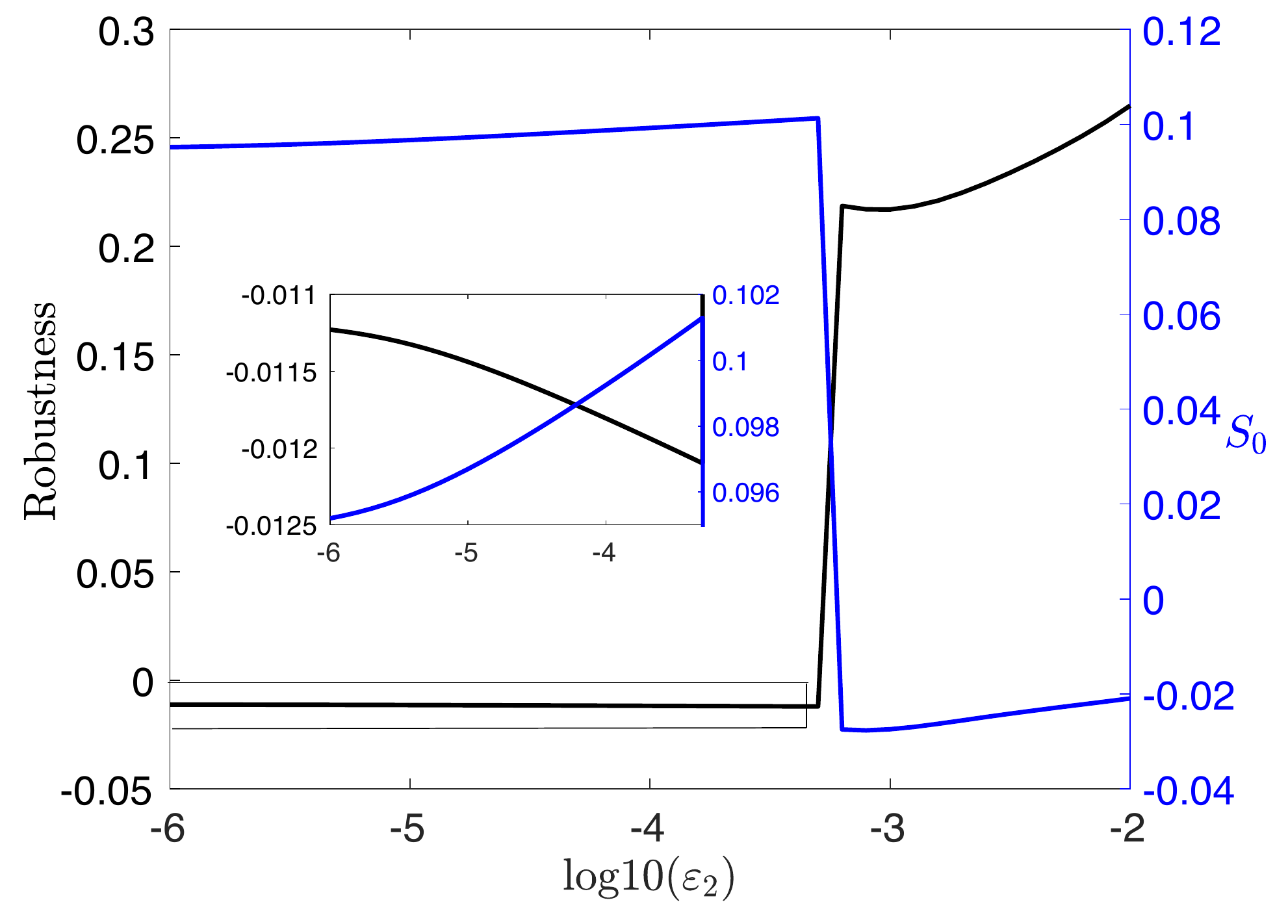}
		\end{center}
		\caption{\label{fig:robustness-eps123} Effects of varying $\varepsilon_0$ (top row), $\varepsilon_1$ (center row) and $\varepsilon_2$ (bottom row) on the robustness \eqref{eq:robustness} to $F_{\text{sw}}$  and the unperturbed seaweed intake rate $S_0$, when  $F_{\text{sw}}=0.01$. Blue curves: Fitness $S_0$.  Black curves: Robustness.   }
	\end{figure}

	Figure \ref{fig:robustness-eps123} shows the seaweed intake rate and robustness to the increased load $F_{\text{sw}}$ with respect to changes in sensory feedback strengths $\varepsilon_i$, $i\in\{0,1,2\}$.
	The performance $S_0$ becomes negative when $\varepsilon_0$ or $\varepsilon_1$ is relatively small (e.g., smaller than $10^{-5}$) or when $\varepsilon_2$ is relatively big (e.g., larger than $10^{-3}$), during which the system is in a fast limit cycle/biting mode and hence cannot swallow seaweed. 
	
	When the system is in the heteroclinic/swallowing mode, as one might expect, increasing the sensory feedback (e.g., $\varepsilon_2$) improves the performance. Surprisingly, our results show that increasing sensory feedback strengths to the two protraction-related neural pools leads to opposite results by decreasing the performance. 
	These results seem to suggest that to restore the deficit caused by the increased load and achieve an increased robustness, we can either increase $\varepsilon_2$ or decrease $\varepsilon_0$ and/or $\varepsilon_1$. However, this is not true. 
	As shown in Figure \ref{fig:robustness-eps123}, a decrease in the robustness can be induced by either 
	decreasing $\varepsilon_0$ or increasing $\varepsilon_2$. 
	Moreover, the robustness is largely insensitive to changes in $\varepsilon_1$, despite the fact that it influences the performance. 
	Understanding these effects on the robustness would require analysis of a second-order variational problem and represents a future direction for understanding neuromodulation.  
	
	
\subsubsection{Varying muscle strengths.}
\begin{figure*}[!t]
\begin{center}
\includegraphics[width=8cm]{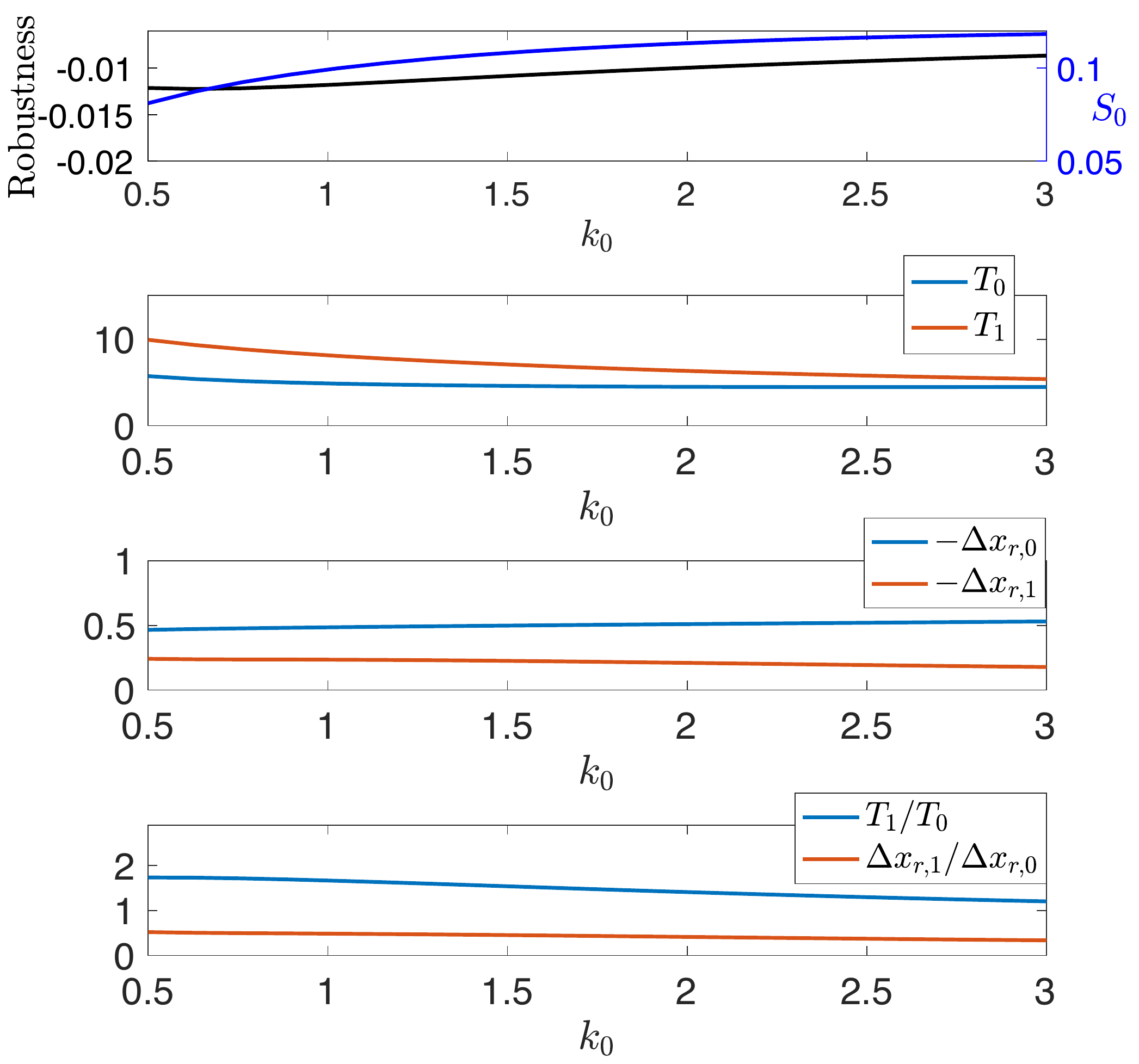}
\includegraphics[width=8cm]{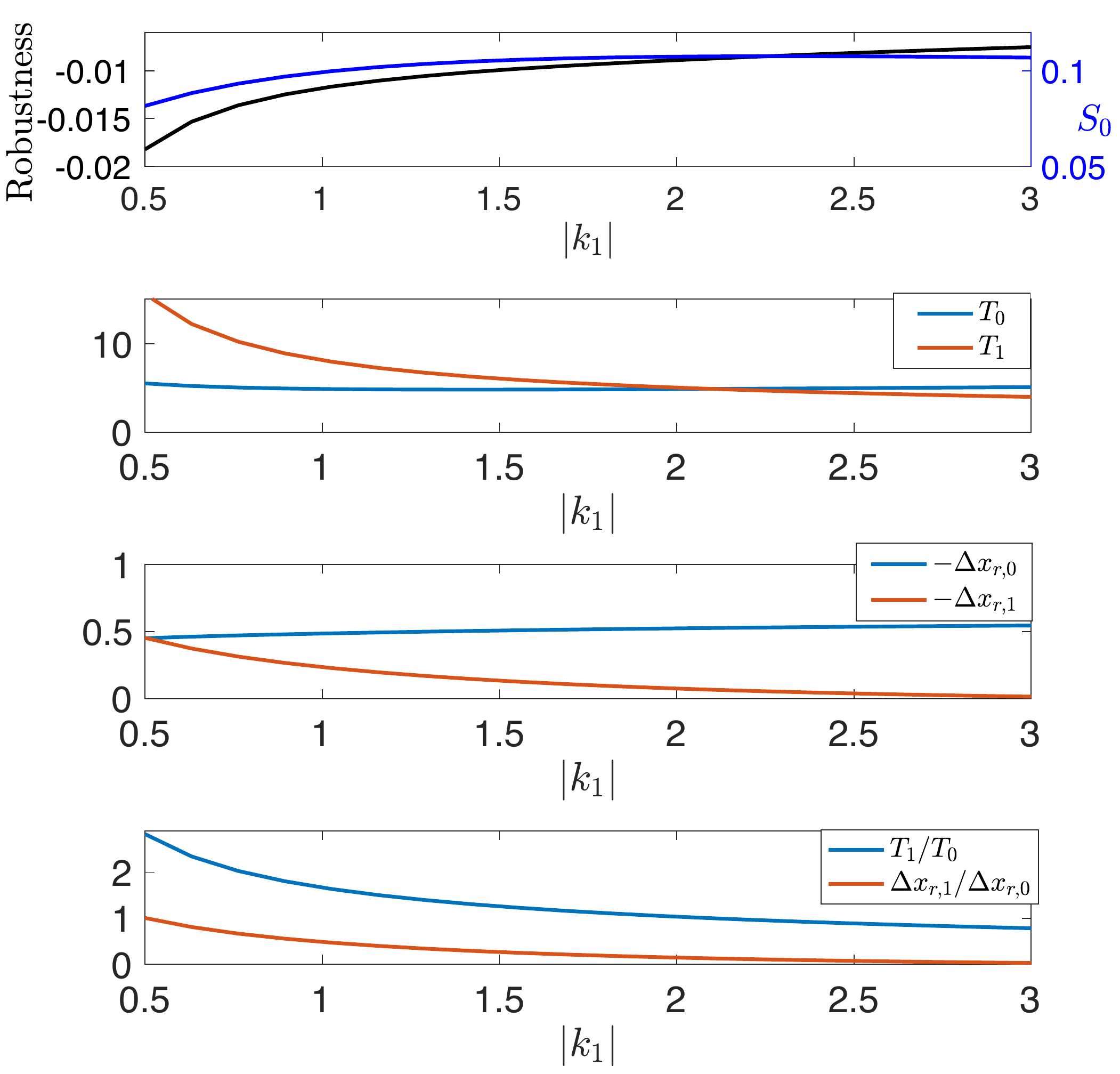}
\end{center}
\caption{\label{fig:neuromodulation_k0_k1} Effects of  varying muscle strengths $k_0$ (left panels) and $|k_1|$ (right panels) on the robustness to $F_{\text{sw}}$ (top panels, black curve) and the unperturbed seaweed intake rate $S_0$ (top panels, blue curve). Default parameters $k_0=1,\,k_1=-1$ represent the strengths and directions of protraction and retraction muscles. The second and third rows of panels show the effects of muscle strengths on timing ($T_0,T_1$) and shape ($-\Delta x_{r,0},-\Delta x_{r,1}$), respectively. The bottom panels shows how  $T_1/T_0$ (blue) and $\Delta x_{r,1}/\Delta x_{r,0}$ (red) change as muscle strengths vary.  
}
\end{figure*} 

Next we investigate how variations of $k_0$ and $|k_1|$, the strengths of the protraction and retraction muscles, affect the robustness to changes in seaweed load. 

Figure \ref{fig:neuromodulation_k0_k1} shows that  performance improves with the increased protractor muscle strength $k_0$ or the increased retractor muscle strength $|k_1|$. This suggests that increasing $k_0$ or $|k_1|$ can help restore the deficit in the performance due to the increased mechanical load and hence boost the robustness, which agrees with our numerical simulations (see Figure \ref{fig:neuromodulation_k0_k1}, top panel, black curve). 

Recall that the robustness can be approximated as $F_\text{sw}\left(\frac{\Delta x_{r,1}}{\Delta x_{r,0}}-\frac{T_1}{T_0}\right)$ (see equation \eqref{eq:robustness}). 
Understanding the underlying mechanisms of the robustness requires one to investigate how the two quantities involving shifts in shape and timing change with respect to $k_0$ or $|k_1|$ (see Figure \ref{fig:neuromodulation_k0_k1}, lower three panels). 
We find that  increasing $k_0$ or $|k_1|$ reduces $T_1$ and $-\Delta x_{r,1}$ while $T_0$ and $-\Delta x_{r,0}$ are almost unaffected. Hence, both $\Delta x_{r,1}/\Delta x_{r,0}$ (the ``stronger" effect in response to perturbations on the seaweed load) and $T_1/T_0$ (the ``longer" effect ) are decreased as we increase the muscle strengths. 
However, the reduction in the ``stronger" effect is smaller than the reduction in the ``longer" effect. 
As a result, the robustness approximated by $F_\text{sw}\left(\frac{\Delta x_{r,1}}{\Delta x_{r,0}} - \frac{T_1}{T_0}\right)$ increases as $k_0$ or $|k_1|$ increases. 

Together, our analytical tools suggest ways in which coordinated changes in intrinsic parameters could maintain fitness and thus enhance robustness.

	\section{Methods}\label{sec:methods}
	
	In this section, we review the classical variational theory for limit cycles (e.g., \citep{filippov1988,bernardo2008,LN2013,park2018}), and new tools that we recently developed in \cite{WGCT2021} for linear approximation of the effects of small sustained perturbations on the timing and shape of a limit cycle trajectory in both smooth and nonsmooth systems. 
	
	In the next two sections we treat the smooth and nonsmooth cases, respectively.  
	In each case, we consider a one-parameter family of $n$-dimensional dynamical systems 
	\begin{equation}\label{eq:dxdt=Feps}
	\frac{d\mbx}{dt}=F_\varepsilon(\mbx),
	\end{equation}
	indexed by a parameter $\varepsilon$ representing a static perturbation of  a reference system 
	\begin{equation}\label{eq:dxdt=F}
	\frac{d\textbf{x}}{dt}=F_0(\textbf{x}).
	\end{equation}
	
\subsection{Timing and shape responses to static perturbations in smooth systems.}
			
Following \cite{WGCT2021}, we make the following assumptions:
	\begin{assumption}\label{ass:smooth} 
	\quad
			\begin{itemize}
				\item The vector field $F_{\varepsilon}(\mbx): \Omega \times \II \to \mathbb{R}^n$ is $C^1$ in both the coordinates $\mbx$ in some open subset $\Omega\subset \mathbb{R}^n$ and the perturbation $\varepsilon\in \II\subset \mathbb{R}$, where $\II$ is an open neighborhood of zero. 
				\item For $\varepsilon\in \II$, system \eqref{eq:dxdt=Feps} has a linearly asymptotically stable limit cycle $\gamma_\varepsilon(t)$, with finite period $T_\varepsilon$ depending  (at least $C^1$) on $\varepsilon$.
			\end{itemize} 
		\end{assumption}
		
		It follows from Assumption \ref{ass:smooth} that when $\varepsilon=0$, $F_0(\mbx)$ is $C^1$ in $\mbx\in \Omega$ and the unperturbed system \eqref{eq:dxdt=F} exhibits a $T_0$-periodic linearly asymptotically stable limit cycle solution $\gamma_0(t) = \gamma_0(t+T_0)$ with $0<T_0<\infty$.
		%
		%
		Assumption \ref{ass:smooth} also implies that the following approximations hold:
			\begin{equation} \label{eq:assump1}
			F_{\varepsilon}(\mbx) = F_0(\mbx) + \varepsilon \frac{\partial F_\varepsilon}{\partial \varepsilon}(\mbx)\Big|_{\varepsilon=0} + O(\varepsilon^2),
			\end{equation}
			\begin{equation}\label{eq:T_eps_expansion} T_\varepsilon=T_0+\epsilon T_1+O(\epsilon^2),
			\end{equation}
			\begin{equation}\label{eq:x_epsilon_of_tau} \gamma_\varepsilon(\tau_\epsilon(t)) = \gamma_0(t)+\varepsilon\gamma_1(t)+O(\varepsilon^2)\quad (\text{uniformly in }t), 
			\end{equation}
			where $T_1$ is the linear shift in the limit cycle period in response to the static perturbation of size $\varepsilon$. 
			This global timing sensitivity, $T_1$, is strictly positive if increasing $\varepsilon$ increases the period. The perturbed time $\tau_\epsilon(t)$ satisfying $\tau_0(t)\equiv t$ and $\tau_\varepsilon(t+T_0)-\tau_\varepsilon(t)=T_\varepsilon$ will be described later; it			allows the approximation \eqref{eq:x_epsilon_of_tau} to be uniform in time and permits us to compare perturbed and unperturbed trajectories at corresponding time points.  
		
		The timing and shape aspects of limit cycles are complementary, and may be studied together by considering the \textit{infinitesimal phase response curve} (IPRC)  and the \textit{variational analysis} of the limit cycle, respectively.  
		
		\paragraph{Infinitesimal Phase Response Curve (IPRC)}
		The IPRC is a classical analytic tool that measures the timing response of an oscillator due to an infinitesimally small perturbation delivered at any given point on the limit cycle. It satisfies the adjoint equation \citep{SL2012}
		\begin{equation}\label{eq:prc}
		\frac{d\z}{dt}=-DF_0(\gamma_0(t))^\intercal \z,
		\end{equation}
		with the normalization condition
		\[
		F_0(\gamma_0(t))\cdot \z(t)=1.
		\]
		The linear shift in period $T_1$ can be calculated using the IPRC as
		\begin{equation}\label{eq:T1}
		T_1=-\int_{0}^{T_0} \z(t)^\intercal\frac{\partial F_\varepsilon(\gamma_0(t))}{\partial \varepsilon}\Big|_{\varepsilon=0}dt.
		\end{equation} 
		
		\ignore{\paragraph{Variational Analysis}
			The variational equation is given by
			\begin{equation}\label{eq:var}
			\frac{du}{dt}=J_0(t)u.
			\end{equation}
			The variational dynamics $u$ represents the linearized dynamics of trajectories near the limit cycle $\gamma(t)$ and describes the time evolution of an instantaneous perturbation on the limit cycle.}

		\paragraph{Forward Variational Equation}
		
		Classical sensitivity analysis \citep{wilkins2009} has been used in many applications to study the shape sensitivity or response of an oscillator to sustained  perturbations. 
		
		The dynamics of the linear shift 
		\begin{equation}
		\label{eq:forward_var_u}
		    \mbu(t)\equiv \lim_{\epsilon\to 0}(\gamma_\epsilon(t)-\gamma_0(t))/\epsilon
		\end{equation} 
		at time $t$ of the periodic orbit $\gamma_\varepsilon(t)$ due to a sustained parametric perturbation $\varepsilon$ initiated at time $0$ satisfies the following forward variational equation:   
		\begin{equation}\label{eq:forward_var}
		\frac{d\mbu}{dt}=DF_0(\gamma_0(t))\mbu+\frac{\partial F_\varepsilon(\gamma_0(t))}{\partial \varepsilon}\Big|_{\varepsilon=0}
		\end{equation}
		with initial condition $\mbu(0)$ set by the difference in the perturbed and unperturbed trajectories at the point where they cross the Poincar\'{e} section defined by the beginning of the closed phase. 
		Specifically, 
		\begin{equation}
		    \label{eq:forward_var_u0}
		    \mbu(0)=\lim_{\epsilon\to 0}(\gamma_\epsilon(t_\epsilon^\text{close})-\gamma_0(t_0^\text{close}))/\epsilon.
		\end{equation}
		Compared with the homogeneous variational equation, which studies the shape sensitivity to instantaneous perturbations, the forward variational equation \eqref{eq:forward_var} contains a non-homogeneous term arising directly from the parametric perturbation acting on the vector field.


		However, since the perturbed limit cycle has a different period $T_{\varepsilon}$ and hence a different perturbed time $\tau_\varepsilon$ due to sustained perturbations, the forward variational equation which neglects such changes in timing fails to give a valid comparison between the perturbed and unperturbed trajectories for times on the order of a full cycle or longer (see Figure \ref{fig:forward_var}C and D). Hence, we adopt a new tool developed in \cite{WGCT2021}, the \textit{infinitesimal shape response curve} (ISRC), which incorporates both the shape and timing aspects and captures a more accurate first-order approximation to the change in shape of the limit cycle under a parameteric perturbation. 
		
		\paragraph{Infinitesimal Shape Response Curve (ISRC)} Suppose the rescaled perturbed time can be written as $\tau_\varepsilon(t)=t/\nu_\varepsilon\in [0, T_\varepsilon]$ for $t\in [0, T_0]$. It follows that the relative change in timing denoted by $\nu_\varepsilon=T_0/T_\varepsilon$ can be represented as $\nu_\varepsilon=1-\varepsilon \nu_1+ O(\varepsilon^2)$ where $\nu_1=\frac{T_1}{T_0}$. 
		
\begin{figure*}[!htp]
\begin{center}
\begin{tabular}{@{}p{0.42\linewidth}@{\quad}p{0.455\linewidth}@{}}
\subfigimg[width=\linewidth]{}{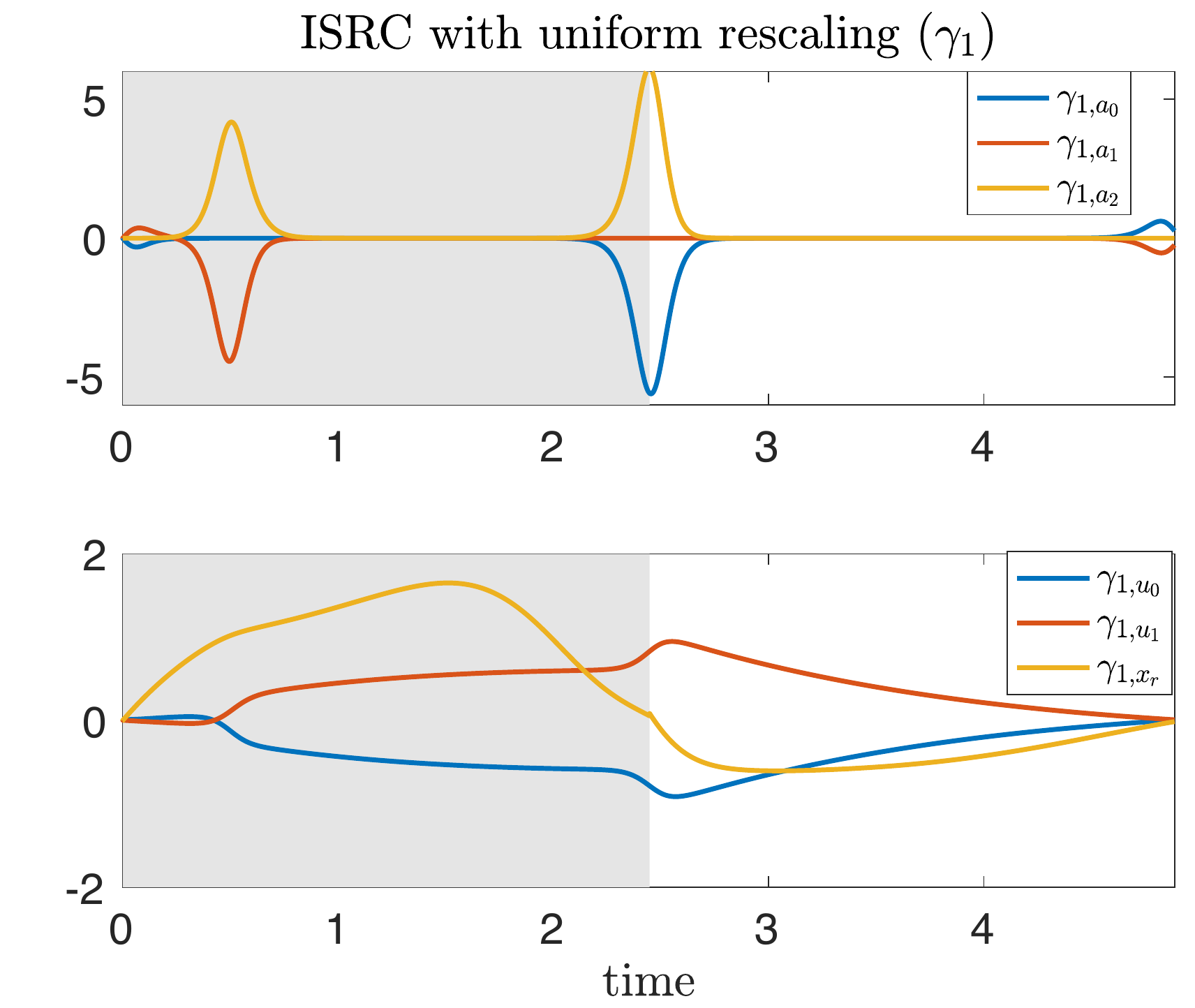}&
\subfigimg[width=0.9\linewidth]{}{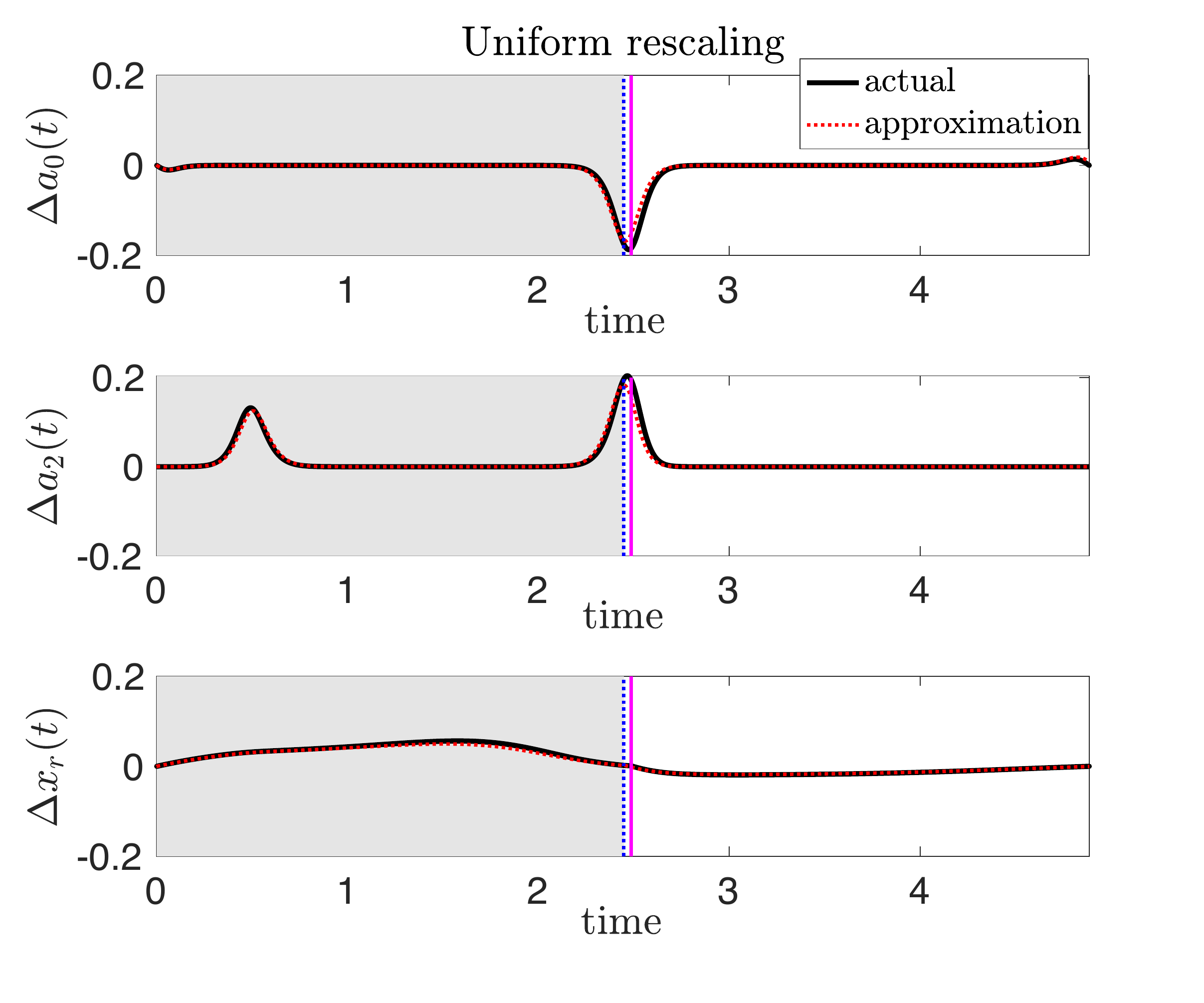}
\end{tabular}
\end{center}
\caption{\label{fig:isrc_open_close} Variational analysis with uniform rescaling. The same perturbation as in Figure \ref{fig:forward_var} is applied to the \textit{Aplysia} model \eqref{eq:Model}. Left: The ISRC $\gamma_1(t)$ with a uniform rescaling over one period. Right: Time series of the difference between the perturbed and unperturbed solutions along $a_0$-, $a_2$-, and $x_r$-directions. The black curve denotes the numerical displacement ($\Delta y(t) = y_\varepsilon(\tau_\varepsilon(t))-y(t)$) computed by subtracting the unperturbed solution trajectory from the perturbed trajectory, after globally rescaling time, and aligning trajectories at the onset of closing. The red dashed curve denotes the product of the perturbation $\varepsilon$ and the ISRC curve. The vertical blue dashed lines indicate the times at which the unperturbed grasper switches from closed to open. Shaded regions and the vertical magenta lines have the same meanings as in Figure \ref{fig:forward_var}. The perturbation is the same as in Figure \ref{fig:forward_var}.}
	\end{figure*}

		\cite{WGCT2021} denote the linear shift in the periodic orbit, $\gamma_1(t)$ in \eqref{eq:x_epsilon_of_tau}, as the ISRC and show it satisfies the following variational equation:
		\begin{equation}\label{eq:isrc}
		\frac{d \gamma_1(t)}{dt}
		= DF_0(\gamma(t)) \gamma_1(t) +\nu_1 F_0(\gamma(t)) +\frac{\partial F_\varepsilon(\gamma(t))}{\partial \varepsilon}\Big|_{\varepsilon=0}.
		\end{equation} 
		This equation resembles the forward variational equation \eqref{eq:forward_var}, but has one additional non-homogeneous term arising from time rescaling $t\to \tau_\varepsilon(t)$. In contrast to the forward variational dynamics $\frac{\partial \gamma_\varepsilon(t)}{\partial \varepsilon}$, the ISRC $\gamma_1(t)$ is periodic with period $T_0$ (see Figure \ref{fig:isrc_open_close}, left). To see how well the ISRC approximates the actual linear shift between the perturbed and unperturbed trajectories, we plot the linear shift approximated from the ISRC (black curve) and the actual displacement (red dashed curve). Overall, they show good agreement with each other except near the transition between the grasper-closed and grasper-open phases. Such discrepancies arise from the fact that the solution segment at the closing phase has different timing sensitivity to the parametric perturbation compared with the segment at the opening phase, as discussed before. While these small errors are nearly unnoticeable (see Figure \ref{fig:isrc_open_close}, right), they expand when the ISRC result is used to calculate the robustness (see Figure \ref{fig:robustness}, bottom panel).    
		
		\ignore{$\z(t)\cdot \frac{\partial \gamma_\varepsilon(t)}{\partial \varepsilon}$ predicts sensitivity of the phase to parametric perturbations, what about $\z(t)\cdot \gamma_1(t)$? Under what circumstances should we use $\gamma_1(t)$ instead of $\frac{\partial \gamma_\varepsilon(t)}{\partial \varepsilon}$ to measure sensitivity?}
		
\begin{figure*}[!htp]
		\begin{center}
	\includegraphics[width=8cm]{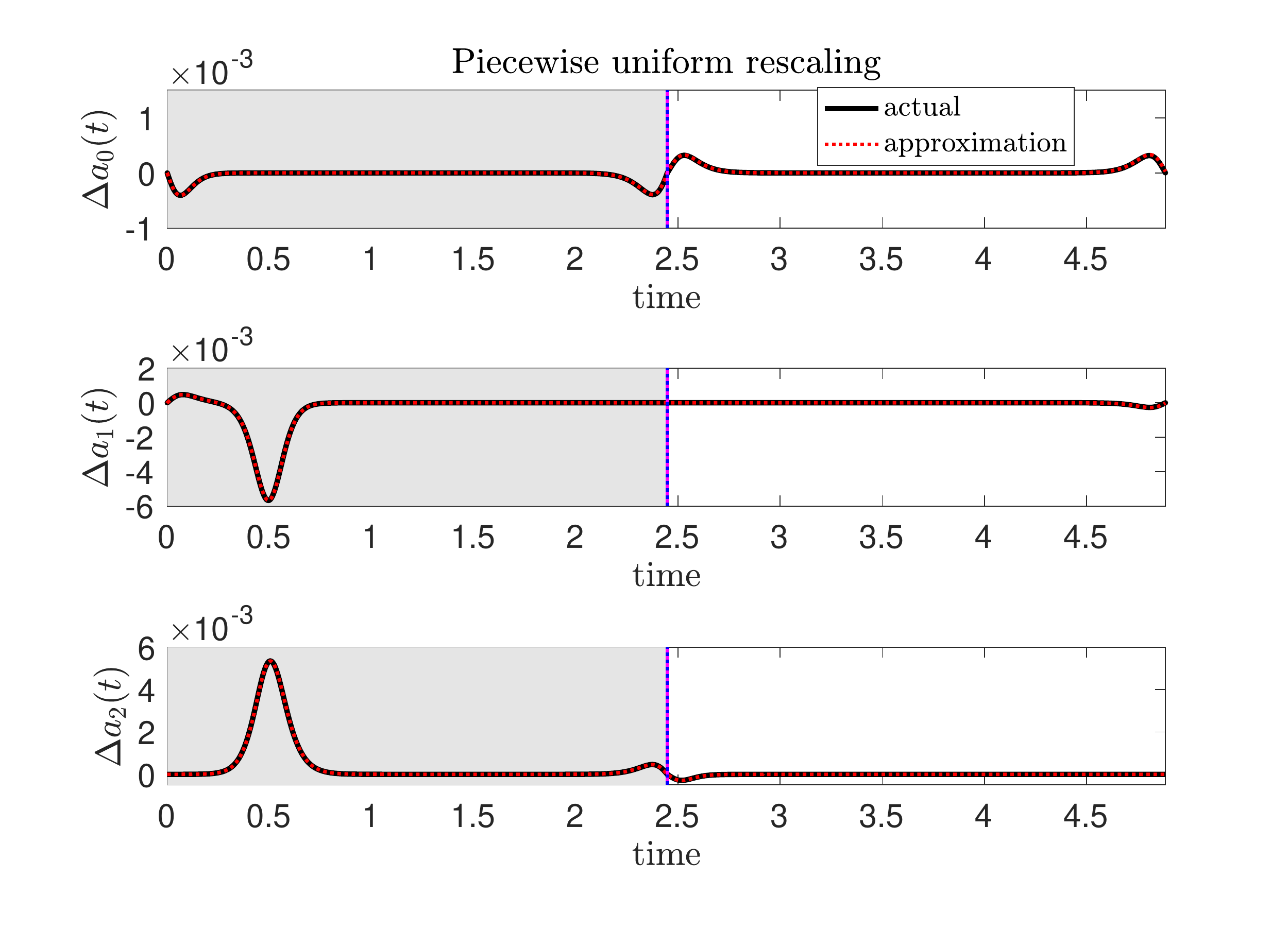}
    \includegraphics[width=8cm]{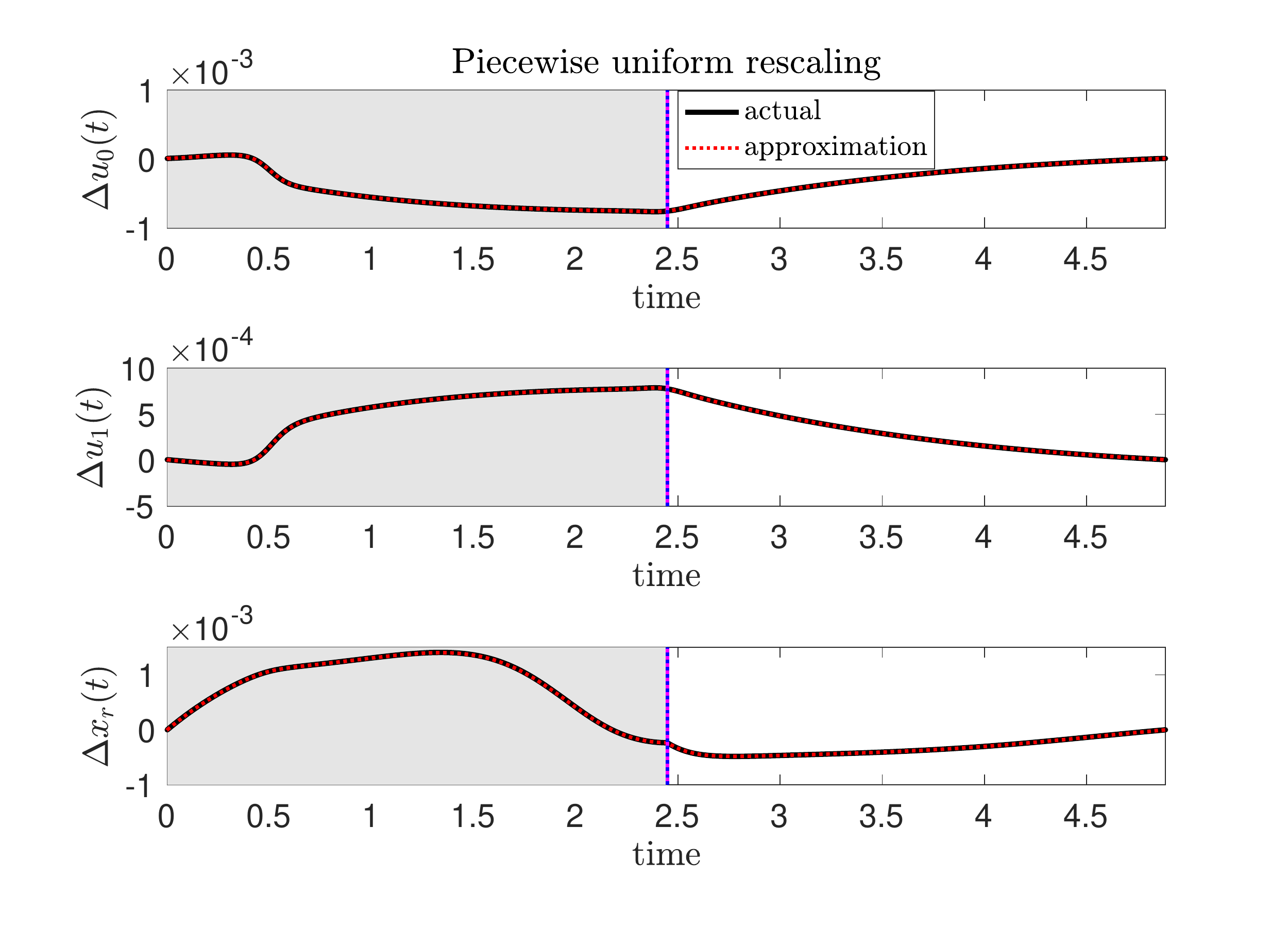}
	\end{center}	\caption{\label{fig:isrc-match}  Displacements between perturbed and unperturbed trajectories estimated from the ISRC $\gamma_1$ with piecewise uniform rescaling (red dashed, $\varepsilon \gamma_1$) agree well with the actual displacement $\Delta y(t) = y_\varepsilon(\tau_\varepsilon(t))-y(t)$ where $y=\{a_0,a_1,a_2,u_0,u_1,x_r\}$ (solid black). Shaded regions, vertical magenta and blue lines have the same meanings as in Figure \ref{fig:isrc_open_close}. The perturbation is the same as in Figure \ref{fig:forward_var}. }
	\end{figure*}

		Thus, in the case when a parametric perturbation leads to different timing sensitivities in different regions, we use the \textit{local timing response curve} (LTRC) defined by \cite{WGCT2021} to compute shifts in timing in different regions in order to improve the accuracy of the ISRC, as demonstrated when considering perturbations to the load applied to the seaweed (see Figure \ref{fig:isrc-match}). 
		
\paragraph{Local Timing Response Curve (LTRC)} The accuracy of the ISRC in approximating the linear change in the limit cycle shape evidently depends on its timing sensitivity, that is, the choice of the relative change in frequency $\nu_1$. In \eqref{eq:isrc}, we chose $\nu_1$ to be the relative change in the full period, by assuming the limit cycle has constant timing sensitivity.
It is natural to expect that different choices of $\nu_1$ will be needed for systems with varying timing sensitivities along the limit cycle. To more accurately capture timing sensitivity of such systems to static perturbations, \cite{WGCT2021} defined a \emph{local timing response curve} (LTRC) which is analogous to the IPRC but measures the linear shift in the time that the trajectory spends within any given region. Specifically, the LTRC is the gradient of the time remaining in a given region until exiting it through some specified Poincar\'{e} section - a local timing surface corresponding to the exit boundary of this region. 
Such a section could be given as a boundary where the dynamics changes between regions, or where a perturbation is applied in one region but not another. For instance, in the feeding system of \textit{Aplysia californica} \citep{shaw2015,lyttle2017}, the open-closed switching boundary of the grasper defines a local timing surface.
		
Let $\eta^{\rm I}$ denote the LTRC vector for region I. Suppose that at time $t^{\rm in}$, the trajectory $\gamma_0(t)$ enters region I upon crossing the surface $\Sigma^{\rm in}$ at the point $\mbx^{\rm in}$; at time $t^{\rm out}$, $\gamma_0(t)$ exits region I upon crossing the surface $\Sigma^{\rm out}$ at the point $\mbx^{\rm out}$. Similar to the IPRC, the LTRC $\eta^{\rm I}$ satisfies the adjoint equation
\begin{eqnarray}\label{eq:LTRC}
\frac{d\eta^{\rm I}}{dt}=-DF(\gamma(t))^\intercal \eta^{\rm I}
\end{eqnarray}
together with the boundary condition at the exit point
\begin{equation}\label{eq:LTRC0}
\eta^{\rm I}(\mbx^{\rm out})=\frac{-n^{\rm out}}{n^{\rm out \intercal} F(\mbx^{\rm out})}
\end{equation}
where $n^{\rm out}$ is a normal vector of $\Sigma^{\rm out}$ at the unperturbed exit point $\mbx^{\rm out}$. The linear shift in the total time spent in region I, $T^{\rm I}_1$, is given by
\begin{equation}\label{eq:local-time-shift}
T^{\rm I}_{1} = \eta^{\rm I}(\mbx^{\rm in})\cdot \frac{\partial \mbx_{\varepsilon}^{\rm in}}{\partial \varepsilon}\Big|_{\varepsilon=0}+\int_{t^{\rm in}}^{t^{\rm out}}\eta^{\rm I}(\gamma(t))\cdot \frac{\partial F_\varepsilon(\gamma(t))}{\partial \varepsilon}\Big|_{\varepsilon=0}dt,
\end{equation} 
where $\mbx_\varepsilon^{\rm in}$ denotes the coordinate of the perturbed entry point into region I. 
It follows that the relative change in frequency local to region I is given by $\nu_1^{\rm I}=T^{\rm I}_{1}/({t^{\rm out}-t^{\rm in}})$.

\paragraph{Piecewise uniform ISRC} The existence of different timing sensitivities of $\gamma(t)$ in different regions therefore leads to a piecewise-specified version of the ISRC \eqref{eq:isrc} with period $T_0$, 
\begin{equation}\label{eq:src-multiplescales}
\frac{d \gamma_1^j(t)}{dt}
= DF_0^j(\gamma(t)) \gamma_1^j(t) +\nu_1^{j} F_0^j(\gamma(t)) +\frac{\partial F_\varepsilon^j(\gamma(t))}{\partial \varepsilon}\Big|_{\varepsilon=0},
\end{equation}
where $\gamma_1^j$, $F_0^j$, $F_{\varepsilon}^j$ and $\nu_1^j$ denote the ISRC, the  unperturbed vector field, the perturbed vector field, and the relative change in frequency in region $j$, respectively, with $j\in\{\rm I, II, III, \cdots\}$.
Note that in a smooth system, $F_0^j\equiv F_0$ for all $j$.

As discussed before, the piecewise-specified ISRC, where $\nu_1$ takes different values in the closing and opening phases, nicely complements the forward variational analysis.
It provides a more self-consistent global description of the shape response of the limit cycle to the mechanical perturbation (see Figure \ref{fig:isrc_pw}). 
Displacements between perturbed and unperturbed trajectories estimated using the piecewise-specified ISRC agree well with the actual displacements (see Figure \ref{fig:isrc-match}). Moreover, it yields a much better approximation to the robustness compared with the ISRC with uniform rescaling (see Figure \ref{fig:robustness}).

		\subsection{Timing and shape responses to static perturbations in nonsmooth systems.}
		
		As discussed before, system \eqref{eq:Model} is a piecewise smooth system with one transversal crossing boundary $\Sigma_{o/c}$ and three hard boundaries ($\Sigma_0,\Sigma_1,\Sigma_2$). The study of limit cycle motions in such nonsmooth systems requires analytical tools beyond the standard arsenal of phase response curves and variational analysis, developed for systems with smooth (differentiable) right-hand sides \citep{spardy2011a,spardy2011b,Park2017}.
		For small instantaneous displacements, variational analysis has been extended to nonsmooth dynamics with both transversal crossing boundaries and hard boundaries for studying the linearized effect on the shape of a trajectory \citep{filippov1988,bernardo2008,LN2013,DL11}. Analysis in terms of infinitesimal phase response curves (IPRC) has likewise been extended to nonsmooth dynamics for studying the linear shift in the timing of a trajectory following a small perturbation, provided the flow is always transverse to any switching surfaces at which nonsmooth transitions occur  \citep{shirasaka2017,park2018,CGL18,wilson2019}. 
		Recently, \cite{WGCT2021} extended the IPRC method to nonsmooth systems with hard boundaries. 
		
	In nonsmooth systems with degree of smoothness one or higher (i.e., \textit{Filippov systems}), the right-hand-side changes discontinuously as one or more switching surfaces are crossed. A trajectory reaching a switching surface or boundary has two behaviors: it may cross the boundary transversally or it may slide along it. Hence, there are two types of boundary crossing points: \textit{transversal crossing points}, at which the trajectory crosses a boundary with finite velocity in the direction normal to the boundary,  and \textit{non-transversal crossing points} including the \textit{landing point} at which a sliding motion along a switching boundary begins, and the \textit{liftoff point} at which the sliding terminates. 
	The time evolutions of the solutions to the variational equation (i.e., the forward variational dynamics and the ISRC) and the solutions to the adjoint equation (i.e., the IPRC and the LTRC) may experience discontinuities at a boundary crossing point \citep{filippov1988,bernardo2008,LN2013,park2018,WGCT2021}.
	
	The discontinuity in the variational dynamics when a trajectory meets a boundary crossing point $\mbx_p$ at crossing time $t_p$ can be expressed with the saltation matrix $S_p$ (see Table \ref{tab:1}): 
		\begin{eqnarray*}
	    {\mbu_p^+=S_p\,\mbu_p^-}
	    \end{eqnarray*}
	where $\mbu(t)$ denotes the solution of the forward variational equation or the ISRC,  $\mbu_p^-=\lim_{t\to t_p^-} \mbu(t)$ and $\mbu_p^+=\lim_{t\to t_p^+} \mbu(t)$ represent the solution just before and just after the crossing, respectively.
	
	The discontinuity in $\z(t)$, the solution to the adjoint equation, at a boundary crossing point $\mbx_p$ may be expressed with the forward jump matrix ($J_p$) \begin{eqnarray}\label{eq:jump-def}
	\z_p^+=J_p\z_p^-
	\end{eqnarray}
	where $\z_p^-=\lim_{t\to t_p^-} \z(t)$ and $\z_p^+=\lim_{t\to t_p^+} \z(t)$ are the IPRC or the LTRC just before and just after crossing the switching boundary at time $t_p$ in forwards time. However, \cite{WGCT2021} showed that the jump matrix is not well defined at a liftoff point and hence introduced a time-reversed version of the jump matrix, denoted as $\mathcal{J}_p$, defined as follows:
	\begin{eqnarray}\label{eq:jump-def-rev}
	\z_p^-=\mathcal{J}_p\z_p^+
	\end{eqnarray}
	Table \ref{tab:1} summarizes the saltation and jump matrices at different types of boundary crossing points.
	
    \begin{table*}[!htp]
\caption{Saltation matrices and jump matrices at boundary crossing points in Filippov systems \citep{filippov1988,bernardo2008,LN2013,park2018,WGCT2021}}
\label{tab:1}       
    \begin{center}
	\begin{tabular}{llll}
	\toprule
	& Landing Point  & Transversal Crossing Point   & Liftoff Point  \\
	\midrule
	Variational dynamics & $S_p=I-n_pn_p^\intercal$ & $S_p=I+\frac{(F_p^+-F_p^-)n_{p}^\intercal}{n_{p}^\intercal F_p^-}$ & $S_p=I$ \\\midrule
	IPRC \& LTRC (forward time)  & $J_p=I$ & $J_p=(S_p^{-1})^\intercal$ &$J_p$ is undefined\\\midrule
	IPRC \& LTRC (time-reversed)  & $\mathcal{J}_p=I$ &$\mathcal{J}_p=J_p^{-1}$ &$\mathcal{J}_p=I-n_pn_p^\intercal$\\\bottomrule
	\end{tabular}
	\end{center}
	In the table, $S_p$, $J_p$ and $\mathcal{J}_p$ denote the saltation matrix, the jump matrix, and the time-reversed jump matrix at some boundary crossing point $\mbx_p=\mbx(t_p)$, respectively. $F_p^-=\lim_{\mbx\to \mbx_p^-}F(\mbx)$ and $F_p^+=\lim_{\mbx\to \mbx_p^+}F(\mbx)$ denote the vector fields of the nonsmooth system just before and just after the crossing at $\mbx_p$, $I$ denotes the identity matrix, $n_p$ denotes the unit normal vector of the crossing boundary at $\mbx_p$. 
	\end{table*}

	\subsection{Simulation codes.} 
	Simulation codes written in Matlab are available at \url{https://github.com/yangyang-wang/AplysiaModel}.
		
		
		
		
		\section{Discussion}\label{sec:discussion}
		
    \paragraph{Overview.}		
		
	Motor systems are robust - they maintain their performance despite perturbations. Understanding the mechanisms of robustness is important but challenging. To unravel the contributions of different components of robustness, we adopted tools we established in \cite{WGCT2021} and reviewed in the methods section (\S\ref{sec:methods}) for studying combined shape and timing responses of both continuous and nonsmooth limit cycle systems under small sustained perturbations. We applied these tools to understand the mechanisms of robustness in a neuromechanical model of triphasic motor patterns in the feeding apparatus of \textit{Aplysia} developed in \citep{shaw2015,lyttle2017}. We show in the results section (\S\ref{sec:results}) that this framework lets us analyze how a small sustained perturbation alters the shape and timing of a closed loop system, and thus we began to describe how the neural and biomechanical components interact to contribute to robustness.
	
	The first perturbation we considered was a sustained increase in mechanical load ($F_{\text{sw}}\to F_{\text{sw}}+\varepsilon$). To our surprise, we discovered that long before sensory feedback affected the system, biomechanics played an essential role in robustness by producing an immediate force increase to resist the applied load (Figure~\ref{fig:forward_var}, \ref{fig:biomechanics} and \ref{fig:forward_var_detail}). Furthermore, although the sensory feedback immediately responded to the perturbation, its effect was delayed by the hard boundary properties of the neural firing rates. Our analysis suggests that sensory feedback contributes to the robustness primarily by shifting the timing of neural activation as opposed to changing neuronal firing rate amplitude (Figure~\ref{fig:isrc_pw}, \ref{fig:prc-aplysia} and \ref{fig:LTRC-close}). Our methods can also be readily used to quantify how changes in timing and shape of trajectory affect the robustness (Figure~\ref{fig:robustness}). We find that sensory feedback and biomechanics contribute to the robustness of the system by generating a stronger retractor muscle force build-up during the prolonged retraction-closed phase that resists the increased load. The increased retractor muscle force ultimately leads to more seaweed being consumed during the slightly longer cycle time despite the large opposing forces, thereby contributing to a robust response. These new insights have refined and expanded a previous hypothesis that sensory feedback is the major mechanism that plays a crucial role in creating robust behavior \citep{lyttle2017}.  
    
    Robustness is sensitive to other model parameters. For example, in \S\ref{ssec:other-param} we investigated how varying internal parameters such as strengths of sensory feedback and muscle activity can help restore the performance that was reduced by an increased applied load (Figure \ref{fig:robustness-eps123} and \ref{fig:neuromodulation_k0_k1}). Again, we obtained some non-intuitive results. For example, increasing the sensory feedback strength can reduce  the robustness rather than improving it (Figure~\ref{fig:robustness-eps123}).
    Moreover, increasing sensory feedback gain has opposite effects on performance and robustness,
    whereas increasing the protractor or retractor muscle strength improves both performance and robustness. Understanding sensitivities of performance to mixed parameters requires us to go beyond our existing methods. This second-order sensitivity represents an interesting future direction for understanding neuromodulation - the coordinated change of multiple system parameters in order to most effectively counter the effect of an external perturbation \citep{Cropper2018}.  There are multiple pathways for neuromodulation, and the simplicity of the model lends itself to detailed analysis of multifactor sensitivities. In future work, we may apply the variational tools used in the present paper for understanding how changes in multiple parameters simultaneously could impact model performance and robustness (cf.~\S\ref{ssec:other-param}).
	
	\paragraph{Experimentally testable predictions.}
    
	The surprising result that the length-tension curves of the opposing muscles generate an instantaneous response to force perturbations could be tested, at least initially, using some of the more realistic biomechanics models that have been developed of \textit{Aplysia} feeding.
	
	For example, in a detailed kinetic model that does not have sensory feedback \citep{Sutton2004, Novakovic2006}, one could apply a step increase in force when the odontophore is closed and the retractor muscle is activated while measuring the force resistance to that change, and compare that to a purely passive response in which the retractor muscle is not activated. The results of this paper predict that there will be significant differences between these conditions.
	
	In a model that does have sensory feedback \citep{WGTC2020}, one could apply a step increase in force when the odontophore is closed and measure the change in force and the duration of the cycle to determine how that perturbation alters fitness. This paper's results predict that the response to a sustained perturbation will be smaller in the presence of sensory feedback and will be larger if sensory feedback is removed.
	
	The model suggests that there may be delays from the time that sensory feedback is available to the time that force changes. Using the model, sinusoidal force changes could be applied at different frequencies to determine the predicted phase lag, and this effect could be tested in the real animal.
	
	Results shown in Figure~\ref{fig:robustness-eps123} suggest that the model is relatively insensitive to changes in the strength of sensory feedback over a wide range of gains. Thus, one experimental test might be to increase or decrease the strength of sensory feedback to show that robustness to changing mechanical loads is not significantly affected. One way to test this hypothesis would be to use the newly developed technology of carbon fiber electrode arrays, which could be used to excite, inhibit, and record from many sensory neurons simultaneously \citep{Huan2021}.
	
	In contrast, results shown in Figure~\ref{fig:neuromodulation_k0_k1} suggest that changing the relative strengths of the muscles can have larger effects on robustness. Previous studies have shown that neuromodulators can speed up and strengthen muscular contractions and thus might contribute to robustness \citep{TaghertNitabach2012, Lu2015, Cropper2018}. Studies of the neuromuscular transform \citep{Brezina2000} suggested that neuromodulation could effectively speed up and strengthen feeding responses in normal animals, and thus might contribute to robustness.
	
	Future experimental studies could be guided by coordinated changes of parameters in this model using the analysis tools we have presented.
		
	\paragraph{Caveats and limitations.}
	    
	Tracking possible transitions into and out of constraint surfaces becomes combinatorially complex as the number of distinct constraint surfaces grows. 
	Here we impose three hard boundaries at $a_i\ge 0$, as discussed above, by requiring firing rates to be nonnegative.  
	An earlier model specification given in \citep{shaw2015, lyttle2017} also required firing rates to be bounded via the constraint
	$a_i\le 1$.  
	Here we relax this constraint for computational convenience, since the coexistence of multiple constraints requires encoding entry/exit conditions and vector field restrictions for all feasible combinations of constraints.  
	In practice, comparison of simulations with and without the $a_i\le 1$ constraint give qualitatively and quantitatively indistinguishable results under most conditions.
	
	Our analysis is in principle limited to small perturbations. Large perturbations lead to crossing of bifurcation boundaries in which the behavior switches to a different dynamical mode. ``Robustness'' in a broader sense can mean the distance to a basin of attraction of another dynamical attractor. For example, if the force is increased too much, the model will collapse into a stable fixed point with overextended protraction, while the animal will engage a different response to release or sever the seaweed to avoid damage to its feeding apparatus.  This aspect is not captured in the variational approach.  Nonlinear and bifurcation analysis could complement the present study and is ripe for investigation in future work.

	In this paper we considered a specific perturbation, namely increasing the force opposing seaweed ingestion $F_\text{sw}\to F_\text{sw}+\epsilon$.  
	Note that in this formulation, the perturbation parameter $\epsilon$ carries the same units (force) as $F_\text{sw}$. 
	Consequently, in order to use a unitless measure of robustness, the expression \eqref{eq:robustness} includes a factor of $F_\text{sw}/\epsilon$.  
	Also, in this formulation, the timing sensitivity $T_1$ (shift in period \emph{per increase in force}) and shape sensitivity $\gamma_1$ (shift in limit cycle shape \emph{per increase in force}) have units including reciprocal force.  
	As an alternative formulation, which might facilitate comparison of robustness to perturbations across different modalities, one could rewrite the force perturbation as $F_\text{sw}\to F_\text{sw}(1+\epsilon)$.  
	In this case $\epsilon$ would represent a unitless measure of \emph{relative} perturbation size.
	The subsequent variational, IPRC, ISRC and LTRC analysis would remain unchanged, except the resulting quantities $Z$, $T_1$, $\gamma_1$, and $\eta_1$ would undergo a change in units, hence a multiplicative (fixed) change in scale.   
	An advantage of specifying perturbations as a relative or unitless quantity would be that a similar analysis to that undertaken in this paper could be applied to other modalities in the same or system or across disparate systems.

    \paragraph{Generalizability to other systems.}

	Although we focused in the present work on the robustness of the mean rate of seaweed intake with respect to increases in the force opposing ingestion, our analysis carries over to other objective functions (e.g.~calories consumed per energy expenditure) as well as other perturbations (e.g.~temperature, which may alter the speed of feeding in \textit{Aplysia}).  
	The variational approach to analyzing robustness should apply to any reasonable (e.g.~smoothly differentiable) objective function and any parameter represented in the system, e.g.~adjustments to changes in speed, steepness, or right-left asymmetry of walking movements on a (split) treadmill system \citep{Frigon2013,Embry2018}.
	
	The present manuscript applies variational methods to understand the robustness in a specific \textit{Aplysia} neuromechanical model \citep{lyttle2017}.
	This model makes significant simplifications to the real feeding apparatus control system in order to gain mathematical tractability and analytical and biological insights. Nonetheless, the framework developed in \citep{WGCT2021} applies naturally to more elaborate dynamical models of \textit{Aplysia} feeding such as \citep{WGTC2020} and models incorporating conductance-based network descriptions of the central pattern generator  \citep{Cataldo2006,Costa2020}.
	Thus, what we have done here provides a framework for understanding neural control of motor behaviors like the one considered in this paper. 
	
	More broadly, motor control beyond the \textit{Aplysia} feeding system is also amenable to the analysis of the sort developed in \S\ref{sec:methods} \citep{WGCT2021}.
	For example, the stability of bipedal walking movements remains a challenge in the field of mobile robotics \citet{Vukobratovic2012,Westervelt2018}. 
	Biologically inspired robotics continues to provide alternative approaches with greater robustness than conventional devices \citep{beer2009,PLI2009,Beer1997,Goldsmith2019}. The variational framework exhibited here applies to these systems as well \citep{Fitzpatrick2020}.
	In the context of any motor control model, the variational analysis we present here should allow analysis of robustness of any reasonable objective function with respect to any system parameter.


\section{Acknowledgments}
This work was supported in part by National Institutes of Health BRAIN Initiative grant RF1 NS118606-01 to HJC and PJT,  by NSF grant DMS-2052109 to PJT, and by NSF grant DBI 2015317 to HJC, as part of the NSF/CIHR/DFG/FRQ/UKRI-MRC Next Generation Networks for Neuroscience Program.
This work was supported in part by the Oberlin College Department of Mathematics.
We thank Zhuojun Yu for providing a critical reading of the manuscript.

%
		


		\appendix
		
		\section{Tables for model parameters and initial conditions}
		
		Values for model parameters and initial conditions of state variables are given in Table \ref{tab:parameters} and Table \ref{tab:initial_conditions}. 
		
        
        \begin{table}[!ht]
        \centering
        \begin{tabular}{|c|c|p{35mm}|}
             \hline
             Parameter & Value & Description \\
             \hline
             $\gamma$ & 2.4 & inhibition strength from next pool \\
             $\epsilon_i$ & $10^{-4}$ & sensory feedback strength \\
             $\mu$ & $10^{-6}$ & neural pool intrinsic excitation \\
             $\tau_{a}$ & 0.05 & neural pool time constant \\
             $\tau_{m}$ & 2.45 & muscle activation time constant \\
             $b_\textrm{r}$ & 0.4 & grasper damping constant\\
             $c_{0}$ & 1.0 & position of shortest length for I2\\
             $c_{1}$ & 1.1 & position of center of I3\\
             $F_{\textrm{sw}}$ & 0.01 & force on the seaweed resisting ingestion \\
             $\sigma_{0}$ & -1 & sign of proproceptive input to $a_0$ motor pool\\
             $\sigma_{1}$ & 1 & sign of proproceptive input to $a_1$ motor pool \\     
             $\sigma_{2}$ & 1 & sign of proproceptive input to $a_2$ motor pool \\
             $\xi_{0}$ & 0.5 & proprioceptive neutral position for protraction-open neural pool\\
             $\xi_{1}$ & 0.5 & proprioceptive neutral position for protraction-closed neural pool\\
             $\xi_{2}$ & 0.25 & proprioceptive neutral position for retraction-closed neural pool\\
             $u_{\textrm{max}}$ & 1.0 & maximum muscle activation \\
             $w_{0}$ & 2 & maximal effective length of I2\\
             $w_{1}$ & 1.1 & maximal effective length of I3\\
             $k_{0}$ & 1 & strength and direction of the protrator muscle\\
             $k_{1}$ & -1 & strength and direction of the retractor muscle\\
             \hline
        \end{tabular}
        \caption{Model parameters}
        \label{tab:parameters}
        \end{table}

        
        \begin{table}
        \centering
        \begin{tabular}{|c|c|>{\raggedright}p{35mm}|}
             \hline
             \parbox{10mm}{\centering State\\variable} & %
                \parbox{10mm}{\centering Initial value} & %
                Description \tabularnewline
             \hline
             $a_{0}$ & $0.9$ & activity of I2 motor pool (non-negative)\tabularnewline
             $a_{1}$ & $0.08355$ & activity of hinge motor pool (non-negative)\tabularnewline
             $a_{2}$ & $0.00003$ & activity of I3 motor pool (non-negative)\tabularnewline
             $u_{0}$ & 0.748 & activity of I2 muscle\tabularnewline
             $u_{1}$ & 0.25 & activity of I3 muscle\tabularnewline
             $x_{r}$ & 0.65 & grasper position (0~is retracted, 1~is protracted)\tabularnewline
             \hline
        \end{tabular}
        \caption{State variables}
        \label{tab:initial_conditions}
        \end{table}

		\section{Different timing sensitivities to muscle perturbations}\label{sec:different-time-sensitivity}
		
\begin{figure*}[!htp]
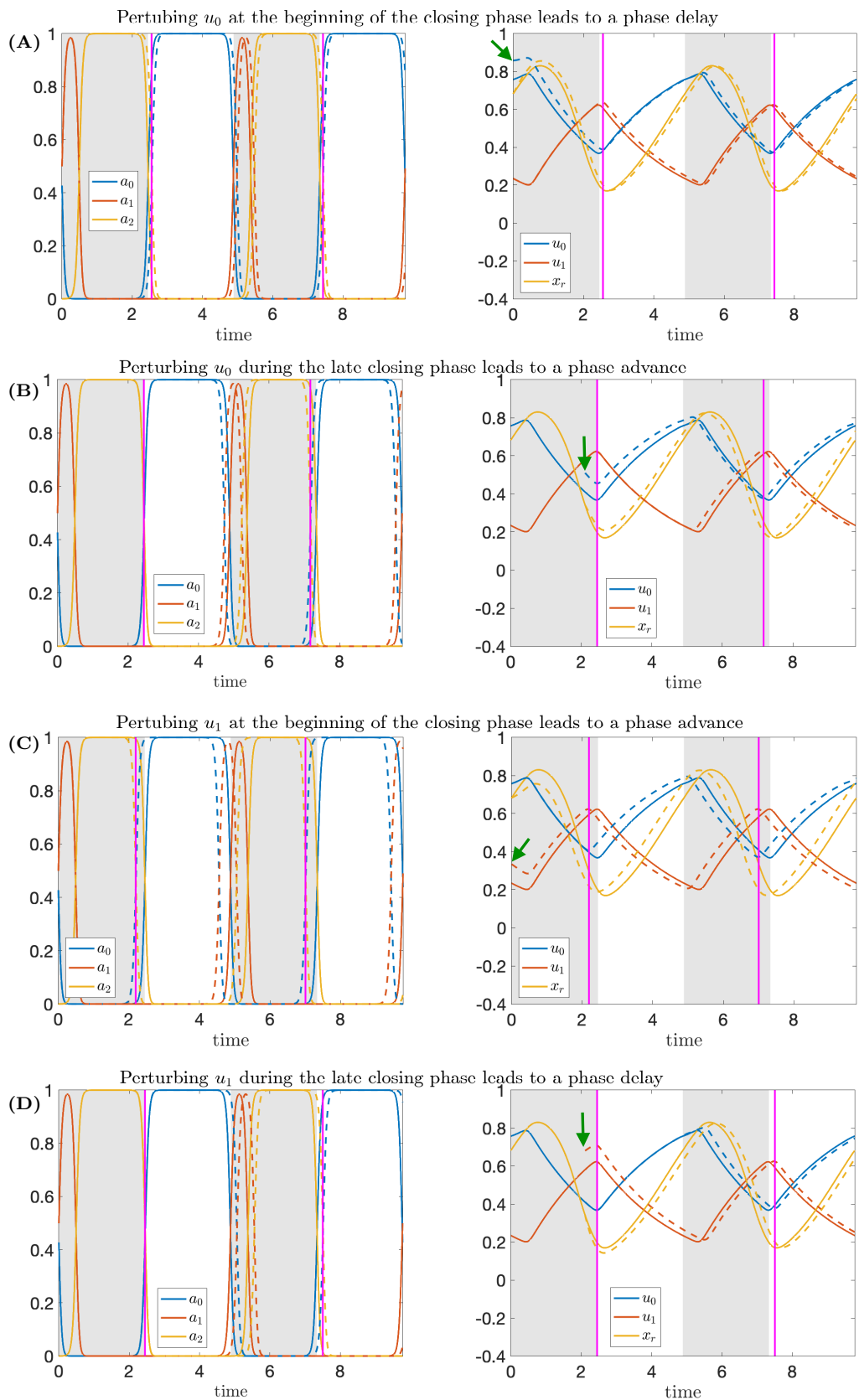

\begin{center}
\begin{tabular}{@{}p{0.8\linewidth}@{}}
\subfigimg[width=\linewidth]{\textbf{{(A)}}}{Fig22.pdf}\\
\subfigimg[width=\linewidth]{\textbf{{(B)}}}{Fig23.pdf}\\
\subfigimg[width=\linewidth]{\textbf{{(C)}}}{Fig24.pdf}\\
\subfigimg[width=\linewidth]{\textbf{{(D)}}}{Fig25.pdf}
\end{tabular}
\end{center}	\caption{\label{fig:iPRC-muscle} Time series of trajectories before (solid) and after (dashed) an instantaneous perturbation of the muscle activation variables ($u_i\to u_i+0.1$, see green arrows). Left panels show trajectories for neural variables, while right panels show trajectories for mechanical variables.  (A) Perturbing the protractor muscle activation $u_0$ at the beginning of the closing phase leads to a phase delay. (B) Perturbing $u_0$ during the late closing phase leads to a phase advance. (C) Perturbing the retractor muscle activation $u_1$ at the beginning of the closing phase leads to a phase advance. (D) Perturbing $u_1$ during the late closing phase leads to a phase delay. Shaded regions and vertical magenta lines have the same meanings as in Figure \ref{fig:forward_var}.}
\end{figure*}
	
Here we explain why increasing the protractor (resp., retractor) muscle activation during the early closing phase leads to a phase delay (resp., phase advance), whereas increasing the muscle activations during the late closing phase lead to the opposite effects (see Figure \ref{fig:prc-aplysia} in \S\ref{sec:results-iprc}). 
		
Early in the closing phase (i.e., the protraction-closed phase), increasing $u_0$ leads to a phase delay. This effect occurs  because with larger $u_0$ force, $x_r$ protracts more, which prolongs the inhibition to $a_0$ through sensory feedback (feedback to $a_0$ is inhibitory when $x_r>0.5$). Hence $a_0$ activates at a later time and the switch from closed to open is delayed, corresponding to a phase delay (see Figure \ref{fig:iPRC-muscle}A). 
		
On the other hand, increasing $u_1$ during the early closing phase leads to a phase advance, because $x_r$ decreases due to the increased retraction muscle forces and hence the inhibition switches to excitation earlier than in the original case (see Figure \ref{fig:iPRC-muscle}C). 

During the late retraction-closed phase, increasing $u_0$ leads to a phase advance (see Figure \ref{fig:iPRC-muscle}B). With increased protractor muscle force, $x_r$ increases, but soon the state transitions to protraction-open. Then, the inhibition on $a_1$ from the sensory feedback (feedback to $a_1$ is inhibitory when $x_r<0.5$) will be released earlier than before, because $x_r$ is larger under perturbation and hence $a_1$ activates earlier. As a result, the system switches from opening to closing phase earlier and this change corresponds to a phase advance. 

On the other hand, if we increase $u_1$ during the late closing phase, a phase delay results because $x_r$ decreases with the perturbation. This effect prolongs the inhibition from sensory feedback to $a_1$, since $x_r$ stays below $0.5$ for a longer time (see Figure \ref{fig:iPRC-muscle}D).


\end{document}